\documentclass[preprint]{aastex}

\usepackage{subfigure,epsfig,longtable,url}

\begin{document}

\title{Sedna and the Oort Cloud Around a Migrating Sun}

\author{Nathan A. Kaib\altaffilmark{1,2,4}, Rok Ro{\v s}kar\altaffilmark{3,4} \& Thomas Quinn\altaffilmark{4}}

\altaffiltext{1}{Department of Physics, Queen's University, Kingston, ON K7L 3N6, Canada, nkaib@astro.queensu.ca}
\altaffiltext{2}{Canadian Institute for Theoretical Astrophysics, University of Toronto, Toronto, ON M5S 3H8, Canada}
\altaffiltext{3}{Institute for Theoretical Physics, University of Z\"{u}rich, Institute for Theoretical Physics, University of Z\"{u}rich}
\altaffiltext{4}{Department of Astronomy, University of Washington, Box 351580, UW, Seattle, WA 98195-1580}

\section{Abstract}
Recent numerical simulations have demonstrated that the Sun's dynamical history within the Milky Way may be much more complex than that suggested by its current low peculiar velocity \citep{selbin02,rosk08b}.  In particular, the Sun may have radially migrated through the galactic disk by up to 5--6 kpc \citep{rosk08b}.  This has important ramifications for the structure of the Oort Cloud, as it means that the solar system may have experienced tidal and stellar perturbations that were significantly different from its current local galactic environment.  To characterize the effects of solar migration within the Milky Way, we use direct numerical simulations to model the formation of an Oort Cloud around stars that end up on solar-type orbits in a galactic-scale simulation of a Milky Way-like disk formation.  Surprisingly, our simulations indicate that Sedna's orbit may belong to the classical Oort Cloud.  Contrary to previous understanding, we show that field star encounters play a pivotal role in setting the Oort Cloud's extreme inner edge, and due to their stochastic nature this inner edge sometimes extends to Sedna's orbit.  The Sun's galactic migration heightens the chance of powerful stellar passages, and Sedna production occurs around $\sim$20--30\% of the solar-like stars we study.  Considering the entire Oort Cloud, we find its median distance depends on the minimum galactocentric distance attained during the Sun's orbital history.  The inner edge also shows a similar dependence but with increased scatter due to the effects of powerful stellar encounters.  Both of these Oort Cloud parameters can vary by an order of magnitude and are usually overestimated by an Oort Cloud formation model that assumes a fixed galactic environment.  In addition, the amount of material trapped in outer Oort Cloud orbits ($a>$ 20,000 AU) can be extremely low and may present difficulties for traditional models of Oort Cloud formation and long-period comet production.


\section{Introduction}
The Oort Cloud is the most remote region of the solar system, and its dynamics are strongly affected by gravitational perturbations beyond the solar system \citep{oort50}.  In particular, it is believed that perturbations from passing stars and the local Galactic tide drive the current orbital evolution of the Oort Cloud \citep{heistre86}.  The strength of both of these perturbers is a function of the density of the local solar neighborhood.  In the current solar environment, the tide of the Milky Way has been shown to be the more powerful perturbation over long timescales \citep{heistre86,mormul86}, and it plays a pivotal role in both enriching and eroding the Oort Cloud population.  This is because the tidal force torques the orbits of distant bodies and causes their perihelia to precess over time \citep{heistre86}.  During Oort Cloud formation, as bodies' orbital semimajor axes are inflated via planetary encounters, this tidal torque can lift these bodies' perihelia out of the planetary region into the Oort Cloud, averting them from being ejected by further planetary encounters \citep{dun87}.  However, once bodies reach the Oort Cloud, the Galactic tide can also lower their perihelia and reinject them into the planetary region where many are eventually ejected to interstellar space by encounters with Jupiter or Saturn \citep{heis87}.

Although the perturbations of passing field stars are weaker than the Galactic tide when averaged over long time periods, they play a fundamental role in Oort Cloud dynamics as well.  Like the Galactic tide, stellar perturbations influence Oort Cloud dynamics mainly by torquing orbital perihelia in and out of the planetary region \citep{oort50}.  However, stellar encounters are a truly stochastic process, and they help keep the Oort Cloud isotropized.  Without their influence, the Galactic tide would quickly deplete large regions of orbital parameter space in the Oort Cloud as bodies are injected into the planetary region and lost through planetary ejections \citep{rick08}.  The isotropizing effect of stellar passages repopulates these regions of orbital space \citep{fouch11}.  Thus, there exists a synergy between the perturbations of field stars and the Galactic tide \citep{rick08, colsar10}.

The first numerical simulations of Oort Cloud formation modeled external perturbations with a fixed Galactic tide and an unchanging population of passing stars \citep{dun87}.  One important result from this work was that only bodies with semimajor axes ($a$) beyond 2000 AU could have their perihelia torqued out of the planetary region into the Oort Cloud, implying that orbits with smaller $a$ would still have perihelia anchored near the planets.  The orbital distribution of observed trans-Neptunian Objects (TNOs) have largely agreed with this result.  However, one glaring exception is the orbit of Sedna \citep{brown04}.  With a perihelion ($q$) of 76 AU, this body no longer interacts with the giant planets, yet its high semimajor axis of $\sim$500 AU indicates that it has been strongly perturbed in the past.   While larger than most TNOs, this semimajor axis is still too small for Sedna's perihelion to have been significantly perturbed by the current local Galactic tide, and Sedna seems to occupy an orbit unexplained by any known dynamical process in the solar system \citep{morblev04}.

Since the seminal work of \citet{dun87}, more recent Oort Cloud formation studies have also included the strong perturbations associated a solar birth cluster \citep{fern97}.  In addition to altering the Oort Cloud's structure and trapping efficiency, these perturbations have been demonstrated to potentially populate the Oort Cloud with orbits near or below Sedna's semimajor axis \citep{fernbrun00, bras06, kaibquinn08}.  Consequently, many studies have interpreted Sedna as a dynamical relic from when the primordial solar system experienced much stronger external perturbations within a dense birth cluster \citep{kenbrom04, morblev04, adams10}.  Alternative mechanisms for Sedna's orbit have also been proposed, including distant interactions with an unseen solar companion \citep{mat05} or perturbations from a Mars-mass body present in the primordial scattered disk \citep{gladchan06}.

An underlying assumption motivating every Sedna origin scenario is that this body cannot belong to the conventional Oort Cloud because perturbations from field stars and the Galactic tide have always been roughly as weak as the present epoch.  However, the modeling of external perturbations in even the most sophisticated models of Oort Cloud evolution remain relatively (and perhaps overly) simple.  They typically view the set of external perturbations experienced by the solar system as essentially a step function with time:  after an early phase of very powerful perturbations from a birth cluster, an unchanging set of weaker perturbing forces due to passing field stars and the Galactic tide continues to sculpt the outer solar system.  In this Paper, we argue that such a treatment is inadequate.  Ignoring the subtleties of the local galactic environment's evolution may skew our understanding of Sedna's dynamical history as well as the bulk properties of the entire Oort Cloud.

\subsection{Solar Migration within the Milky Way}
The strength of perturbations from passing field stars and the Galactic tide has been assumed to be constant because Galactic disk stars in general are thought to largely remain near their birth radii, modulo epicyclic oscillations which, in the current solar neighborhood, are restricted to $\sim$1.5 kpc \citep{bin07}.  Further, the Sun's peculiar velocity (and by extension its epicyclic oscillation) is very small implying that is has not experienced many orbital perturbations in the past.  With a nearly fixed galactocentric distance and small vertical oscillations, the density of the local solar neighborhood (which controls the strength of the Galactic tide and rate of stellar passages) should change very little.  Indeed, when the Sun is integrated in an analytical axisymmetric approximation of the Milky Way's potential it is found to stay within $\sim$100 pc of the midplane and between 8--9 kpc from the Galactic center \citep{mat95}.

The assumption of a time-invariant axisymmetric potential may not be valid, however.  We know that the structure of the Milky Way's disk contains irregular non-axisymmetric components such as the spiral arms.  In addition, the Sun has orbited the Galactic center for a substantial fraction of the Milky Way's history, and the galaxy should have accreted significant amounts of new stars and gas during this time.  \citet{selbin02} found that stars orbiting near the corotation resonance of the transient spiral arms could experience significant changes in their angular momentum.  As a result, many stars had their mean orbital distances significantly changed, yet their peculiar velocities were not substantially increased in many cases. Furthermore, it was found that stars on nearly circular orbits (low peculiar velocities) were more prone to this type of dynamical evolution since their angular orbital velocities remain nearly fixed, allowing them to undergo resonant interactions with the arms for longer periods.

In idealized simulations of Milky Way-like disk galaxy formation, \citet{rosk08b} similarly found significant radial migration akin to that described in \citet{selbin02}. Additionally, \citet{rosk08b} found that over 50\% of stars occupying the solar neighborhood (roughly defined as being between 7-9 kpc) were born elsewhere in the disk, mainly in the interior. The migration process in their simulations resembled a random walk rather than a smooth outward evolution, and consequently many stars occupying the solar neighborhood migrated inward before migrating to their present-day position.  Although the migration takes place at the corotation resonance of a spiral, spirals with different pattern speeds exist in the disk at all times in this simulation, allowing the entire disk to be affected (Roskar {\it et al.} 2011 in prep).  

The results of \citet{selbin02} and \citet{rosk08b} raise the possibility that the Sun's galactocentric distance has not remained fixed throughout the history of the solar system, despite its present-day circular orbit about the Galactic center.  This would help explain a conundrum concerning the Sun's metallicity.  The metallicity of the Sun is 0.14 dex larger than the mean metallicity of other nearby solar-age stars \citep{wiel96, nord04}.  If the Sun formed closer to the Galactic center where the ISM is more metal-rich and then later migrated outward to its current position, then such a result would be expected.  In addition, the process of stellar radial migration may explain why the age-metallicity relation for the solar neighborhood is so flat and contains a great deal of scatter \citep{nord04}.  In a closed-box model for the solar neighborhood, one would expect metallicity to be tightly correlated with stellar age since older stars form from gas that is less enriched by supernovae than younger stars.  However, as radial migration brings stars into the solar neighborhood that formed in different parts of the disk with different ISM metallicities, the tight correlation is destroyed.  Thus, radial stellar migration seems to be consistent with metallicity distributions of the solar neighborhood \citep{selbin02, rosk08b, schobin09, hay08, loeb11}.

\subsection{An Evolving Solar Neighborhood}
If the Sun has migrated within the Milky Way since its birth cluster dispersed, then the solar system would have been exposed to galactic environments (and perturbations) that were markedly different from the local one for some or most of its history.  As one moves closer to the Galactic center, the strength of the Galactic tide will increase as the disk density increases.  In addition, both the typical stellar encounter velocity and the rate of encounters will increase with time.  This may impact our understanding of Sedna's history.  Previous studies that have ruled out Sedna from membership in the classical Oort Cloud have not considered these possible variations in stellar and galactic perturbations.  Moreover, because of the fundamental role that these perturbations play in populating and eroding the entire Oort Cloud, this has the potential to alter the entire distribution of comets throughout the Oort Cloud as well as the number of bodies trapped in this reservoir.  

It is not clear how solar migration would affect the evolution of the Oort Cloud, as few previous works have addressed this possibility.  While \citet{mat95} investigated the effect of a changing tidal strength on the Oort Cloud, these fluctuations were due to the small epicyclic oscillations of the Sun about a circular galactic orbit.  These variations in the Sun's galactic position are much smaller than those expected due to solar migration.  Even so, they found that the long-period comet (LPC) flux near Earth would vary by a factor of $\sim$4 as the tide strength changed.  This was primarily due to the vertical oscillations above and below the disk midplane.  While \citet{mat95} focused on the vertical excursions of the Sun, substantial radial migration of the Sun may have equal or larger consequences for the Oort Cloud since the disk density falls off exponentially with distance from the Galactic center.  

To our knowledge, only the works of \citet{tre93} and \citet{bras10} have explored how the structure of the Oort Cloud may vary if the Sun were located in different regions of the galaxy.  \citet{tre93} employed an analytical treatment of Oort Cloud formation to conclude that a Sun-like star in the galactic halo would have an Oort Cloud that is more extended than our own by a factor of $\sim$2.  In contrast, \citet{bras10} used a computational approach to determine how a drastically different galactocentric distance would impact Oort Cloud formation.  They used an analytical approximation of the Milky Way's potential and numerically modeled Oort Cloud formation within it, assuming various distances to the Galactic center between 2 and 20 kpc.  These distances were held constant for the duration of each simulation.  From this work, it was concluded that shifting the galactocentric distance of the Sun does not affect the total number of bodies that are trapped within the Oort Cloud, but it can shift the location of its inner edge by over an order of magnitude.  However, they found that even their most compact Oort Cloud still had an inner edge beyond Sedna's orbit.

Although \citet{bras10} shows that varying the Sun's galactocentric distance will affect the Oort Cloud's structure, it is not clear how these results apply to a Sun migrating within an evolving galactic disk.  A star undergoing radial migration will not have a fixed galactocentric distance, but instead will sample a large range of different galactic environments throughout the course of its life.  Furthermore, it is not obvious what is the realistic range of possible galactocentric distances for a star that attains a solar type orbit.  Lastly, as mentioned previously, the true potential of the galaxy is only axisymmetric to first order and has undoubtedly evolved throughout the solar system's history.  

Here we present work that uses a purely numerical approach to study Oort Cloud formation around a migrating Sun.  To investigate the effect of radial stellar migration, we use the same Milky Way analog simulation from \citet{rosk08b} and combine it with numerical models of Oort Cloud formation.  Our work is described in the following sections.  In Section \ref{sec:galmeth} we describe our numerical methods.  This includes how we select probable galactic histories for the Sun, how the corresponding galactic tidal histories are calculated, and how we model actual Oort Cloud formation for each history.  Following this, we describe our simulation results in Section \ref{sec:galres}, which is divided into two subsections: the first focuses on the dynamics of Sedna-like orbits, and the second studies the properties of the entire Oort Cloud, paying particular attention to the radial distribution of objects, the trapping efficiency of each Oort Cloud model, and the erosion of the outer Oort Cloud.  Finally, we summarize the conclusions of our work in Section \ref{sec:galcon}.

\section{Numerical Methods}
\label{sec:galmeth}

We attempt to directly combine a galactic simulation with another performed on the planetary scale.  Because of the large dynamical range, the numerical method we employ is fairly complex and consists of three distinct sets of simulations.  The first is a galactic-scale simulation modeling the evolution of a Milky Way analog galaxy \citep{rosk08b}.  In the first subsection, we summarize the galaxy simulation we use and how we select solar analogs from this simulation.  We also provide a detailed discussion of how the tidal field around each solar analog is measured.  The second set of simulations we use are planetary-scale N-body simulations employing SCATR \citep{kaib11a} to model the formation of an Oort Cloud around each of our solar analogs (as well as 3 control cases).  Our second methods subsection describes this set of simulations.   Finally, our last subsection describes a set of Monte Carlo simulations using the impulse approximation to model perturbations from passing field stars on Sedna-like orbits.  

\subsection{Galaxy Simulation}

This galaxy simulation has been used previously to study stellar dynamics \citep{rosk08b} and was run with the $N$-Body + smooth-particle-hydrodynamics (SPH) code GASOLINE \citep{wad04}.  This particular simulation is designed to model the last 10 Gyrs of evolution of the Milky Way following its last major merger with another galaxy.  The initial conditions begin with 10$^6$ dark matter particles and 10$^6$ gas particles.  The dark matter particles are arranged in a spherical Navarro-Frenk-White (NFW) profile \citep{nav97}, and a halo of gas particles in hydrostatic equilibrium is imbedded with the same profile.  The entire system is given a cosmologically motivated spin.  The total mass of the system is chosen to be 10$^{12}$ M$_{\sun}$ with baryons accounting for 10\% of this mass.  The mass budgets for each matter component are split evenly for each particle, giving an initial mass of 10$^5$ M$_{\sun}$ for each gas particle.  As the gas collapses and reaches higher densities, gas particles form star particles whose masses are a fraction of the original gas particle, typically around 3 x 10$^4$ M$_{\sun}$.  The star formation recipe depends on the temperature and density of the local gas and also includes supernova feedback cycles \citep{stin06}.  At $t=10$ Gyrs, the simulated galaxy contains roughly 2.5 x 10$^6$ star particles.  As the gas continues to collapse into a rotating disk, transient spiral arms form spontaneously and cause a radial redistribution of stellar matter \citep{selbin02}.  During this evolution, gravitational forces are fully resolved down to distances of 50 pc, below which they are softened with a spline function \citep{stad01}.  It is important to note that our galaxy is modeled in isolation and consequently does not include full cosmological effects.  However, standard cosmological simulations predict that cosmological accretion and mergers are most important at epochs prior to the last major merger of the Milky Way.   Hence, our model represents the rebuilding of the Milky Way disk after the hierarchical building of the galaxy culminating with the last major merger.

We use our galaxy simulation to approximate the dynamics Sun-like stars may have experienced during the past 4 Gyrs in the Milky Way.  Therefore, we compare key properties of the simulated galaxy and the Milky Way in Table \ref{tab:gal}, and a more detailed comparison can be found in \citet{loeb11}.  As can be seen from Table \ref{tab:gal}, the only property of our simulated galaxy that differs appreciably from the Milky Way is the stellar velocity dispersion of the disk.  Our simulated galaxy has a dispersion that is roughly 50\% greater than the Milky Way.  Because this is a gauge of how eccentric the typical stellar orbit is, it may seem that our galaxy could be overestimating the degree of stellar radial migration compared to the Milky Way.  We note, however, that the key ingredient for the processes described in this paper is the presence of transient spiral arms. The asymmetric structure in our simulation is consistent with observations of external systems \citep{rixzar95}, with m=2 Fourier amplitudes in the range of 0.1-0.3. Further, although the velocity dispersions of old stars are somewhat higher in our models, the heating rates are in fact somewhat slower than those in the Milky Way \citep{holm09}, and follow power laws of $\sigma \sim t^{\sim0.3}$. Should we have unrealistic asymmetries in the disk, we might expect that the heating rates would also be uncharacteristically high, but this is not the case. The higher velocity dispersion is likely mostly a consequence of the fact that the disk in our simulation is slightly more massive than the Milky Way.  We are therefore confident that the degree of orbital evolution present in our models is a reasonable proxy for the history of the Milky Way disk.  We have an extensive paper discussing the details of spiral structure and radial migration in our models in preparation.

\subsubsection{Solar Orbital Histories}
\begin{figure}[tbp]
\centering
\includegraphics[scale=.87]{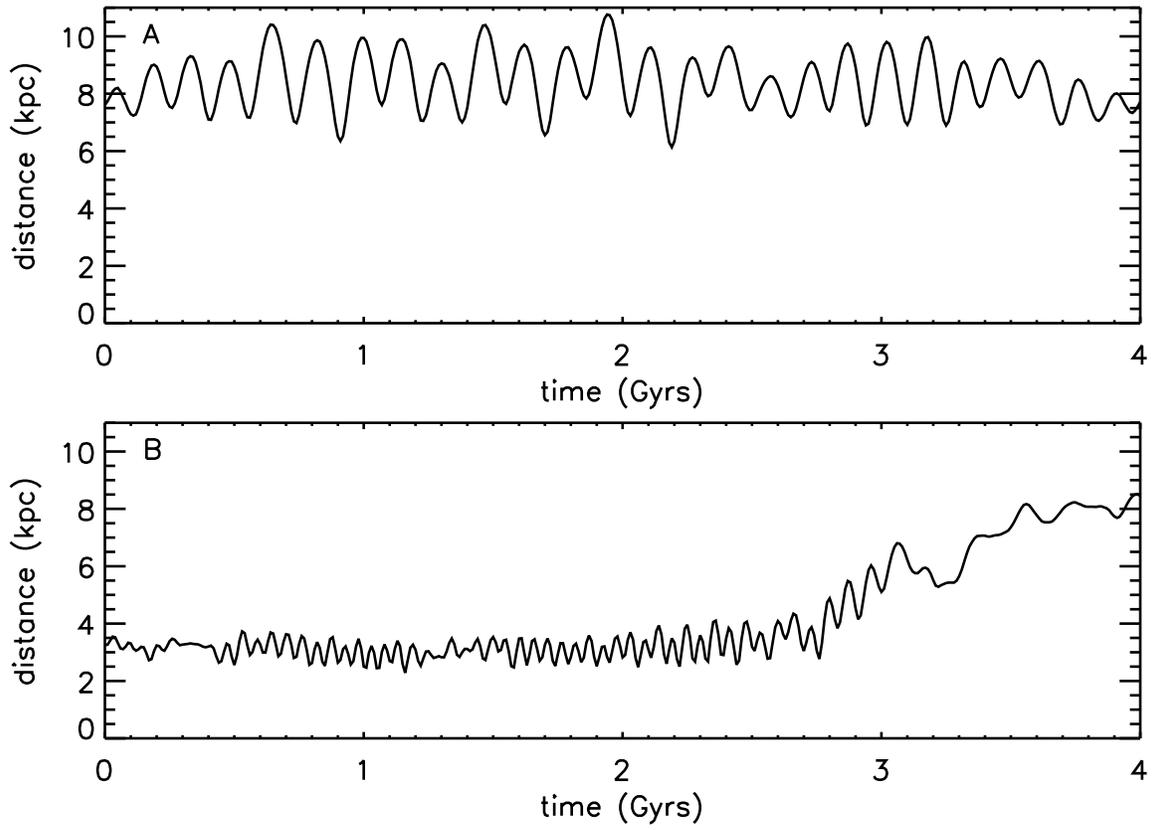}
\caption{Galactocentric distance vs. time for the orbits of two solar analogs.  One is consistent with an approximately fixed galactocentric distance for the history of the solar system ({\it a}), while the other forms much close to the Galactic center than the Sun's current position ({\it b}).}\label{fig:examp}
\end{figure}

To explore the different possible orbital histories of the Sun, we choose solar analog stars found at the last timestep of our galaxy simulation.  Our criteria for choosing solar analogs are based on the stellar age, position, and kinematics.  These criteria are displayed in Table \ref{tab:sun} along with the corresponding solar values for each category.  In total, we found 31 stars that met all of the criteria shown in Table \ref{tab:sun}.  Our suite of solar analogs displays a diverse variety of orbital histories.  For instance, the star particle shown in Figure \ref{fig:examp}a supports the standard idea of the Sun's orbital history, maintaining a roughly constant galactocentric distance for the history of the solar system.  In contrast, Figure \ref{fig:examp}b shows a very different dynamical evolution of a star that also attains a final orbit similar to the Sun's.  This star spends most of its first 3.5 Gyrs orbiting between 2.5--3.5 kpc from the Galactic center before finally migrating outward to 8 kpc in the last Gyr of its history.  

\begin{figure}[tbp]
\centering
\includegraphics[scale=.87]{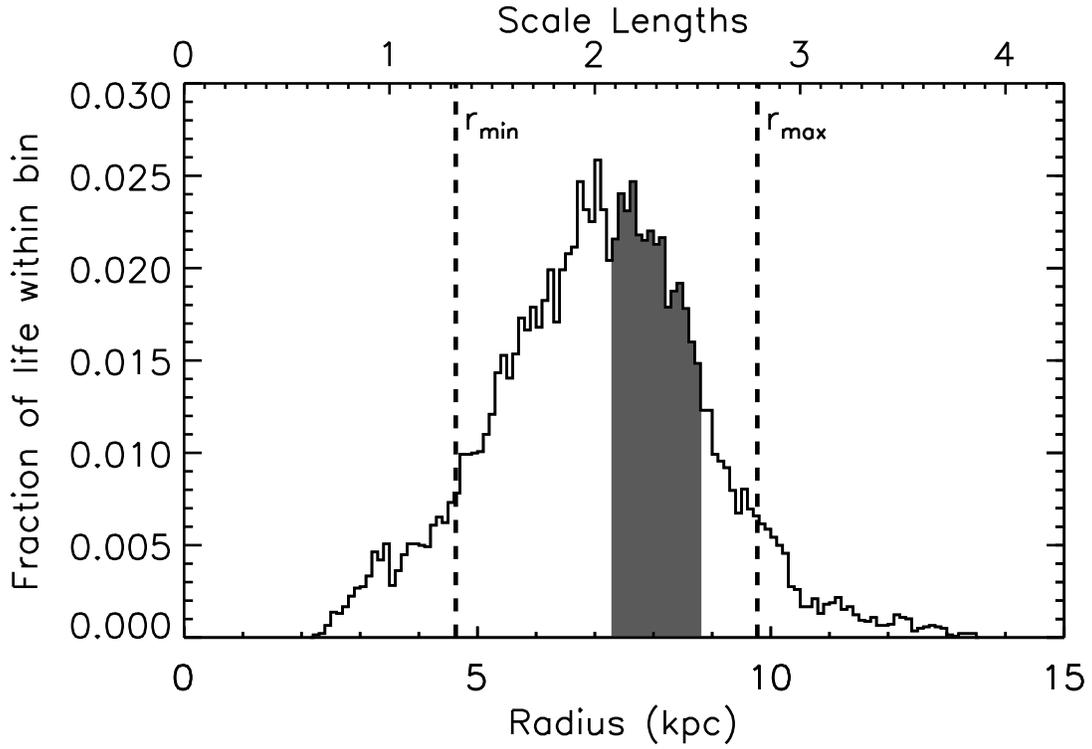}
\caption{Histogram showing the relative time spent at different galactocentric distances for our solar analogs.  The shaded region mark the range of final distances from which our solar analogs were selected.  The dashed lines mark the median values of the minimum and maximum galactocentric distances attained for our solar analogs.  (We assume a scale length of 3.5 kpc in the top x-axis.)}\label{fig:radhist}
\end{figure}

To gauge the full spectrum of possible solar orbital evolutions, we sample the radial distances of each of our solar analog stars every 10 Myrs for the last 4 Gyrs of the galaxy simulation.  This radial data is displayed in the histogram shown in Figure \ref{fig:radhist}.  As can be seen, orbital histories that take a star as close as 2 kpc and as far as 13 kpc from the Galactic center are consistent with the Sun's current position and velocity in our simulated galaxy.  In addition, we also find that the median radial migration incurred during the 4 Gyr orbital histories of our solar analogs is 5 kpc, showing that migration well over a disk scale length is common for Sun-like stars in this Milky Way analog.  This is further demonstrated in Figure \ref{fig:radhist} where we mark the median minimum and median maximum galactocentric distances of our star sample at 4.63 and 9.77 kpc, respectively.  Lastly, the median galactocentric distance of our entire set of stellar radial data is 7.14 kpc, indicating that on average stars with similar dynamical properties to the Sun spend 50\% of their first 4 Gyrs orbiting inside $\sim$7 kpc.

It should be noted that in addition to significant radial migration, our solar analogs can also orbit at different vertical positions above and below the midplane.  However, the effects of this orbital variation on Oort Cloud structure are less significant than radial migration.  This is due to a couple of factors.  First, we have just seen that a typical Sun-like star radially migrates by 5 kpc in our simulation, which is roughly 2 disk scale lengths.  On the other hand,  although there is some vertical variation, our solar analogs tend to stay concentrated toward the midplane for the most part.  99\% of all recorded stellar positions are within 0.5 kpc of the midplane, and 92\% are within 0.3 kpc (one disk scale height).  Thus, the typical vertical variations of our solar analogs modulate the local density by a smaller factor than the typical radial variations.  Furthermore, even if a star drifts far from the midplane it must still make passages through the high density area of the midplane every orbit as it oscillates above and below.  This further diminishes the effect of vertical excitation.

\subsubsection{Tide Approximation}

Developing a method to approximate the Galactic tide near a solar analog is challenging because the potential of our simulated galaxy is not the well-behaved static function employed in many previous Oort Cloud studies.  For instance, it is normally assumed that the vertical component of the Galactic tide dominates over the radial tide.  However, when a spiral arm or any other large substructure is located near one of our stars, this assumption may not be true.  For this reason, we build our tidal field directly from the accelerations calculated by our galactic simulation code.  To do this, we remove the star of interest and place massless test particles offset from the star's position by $\pm$10 pc in the ${\bf \hat{x}}$, ${\bf \hat{y}}$, and ${\bf \hat{z}}$ directions.  We then repeat the force calculations for the timestep following each simulation data output and record the accelerations of each test particle.  Comparing the acceleration data for the test particles then allows us to measure how the acceleration due to the galactic potential varies near the star in the ${\bf \hat{x}}$, ${\bf \hat{y}}$, and ${\bf \hat{z}}$ directions.  (The ${\bf \hat{x}}$, ${\bf \hat{y}}$, and ${\bf \hat{z}}$ vectors define a static coordinate system with an x-y plane in the galactic mid-plane and with vector directions that match the coordinate system chosen in our Oort Cloud formation simulations.)

One potential issue with our tidal calculation technique concerns the coarse mass resolution of galaxy simulations.  In this particular simulation, the masses of stars are scaled up by a factor of $\sim10^5$.  Because the stellar populations of galaxies are collisionless, their dynamics should still be well-modeled by such particles.  However, we are measuring the force variations in this population's potential over scales of tens of pc.  In the real Milky Way, the smooth spatial distribution of gas and millions of stars will produce these force fluctuations, but in our simulated galaxy, it will be a much smaller number of more massive particles.  Because of this, our tidal field measurements can become dominated by a single massive body near the Sun.  To guard against this, the gravitational potential of each particle is softened over a distance $\epsilon$.  This essentially smears the mass of a star particle out over $\epsilon$, so if our test particle configuration happens to lie right next to a 3 x 10$^4$ M$_{\sun}$ star particle, force softening will produce very little acceleration change due to this individual particle's presence.  The characteristic softening length we choose for our tidal calculations is $\epsilon = 150$ pc, which is three times larger than the softening used in the actual galaxy simulation.  We found that without such a large softening length our calculated tidal fields did not behave in a manner expected for a potential that still resembles a smooth disk to first order.  Moreover, it is not surprising that our tidal calculations required a larger $\epsilon$ than the actual force calculations in the original simulation run, since the tidal terms represent an additional spatial derivative with respect to the force terms.  Because the softening length is approximately equal to the scale height of the disk, we should still be able to resolve the actual decrease in stellar density with height above the disk.  In this way, our tidal acceleration measurements become more sensitive to the distribution of many distant star particles rather than a few very close ones.  While this scheme is not perfect, we emphasize that this work is the first of its kind and extrapolating a parsec-scale tidal field from a model of a kiloparsec-scale system requires some degree of compromise.  

Although we expect our tidal field to display time-varying irregularities, our simulated galaxy does resemble a smooth uniform disk to first order.  Consequently, we expect our tidal field to generally reflect the properties of such an idealized galaxy {\it most} of the time.  We use this fact to perform several exercises to verify that our measured tidal field behaves like that of a disk galaxy and is not dominated by shot noise effects from the coarse mass resolution of our Milky Way analog.

\begin{figure}[tbp]
\centering
\includegraphics[scale=.87]{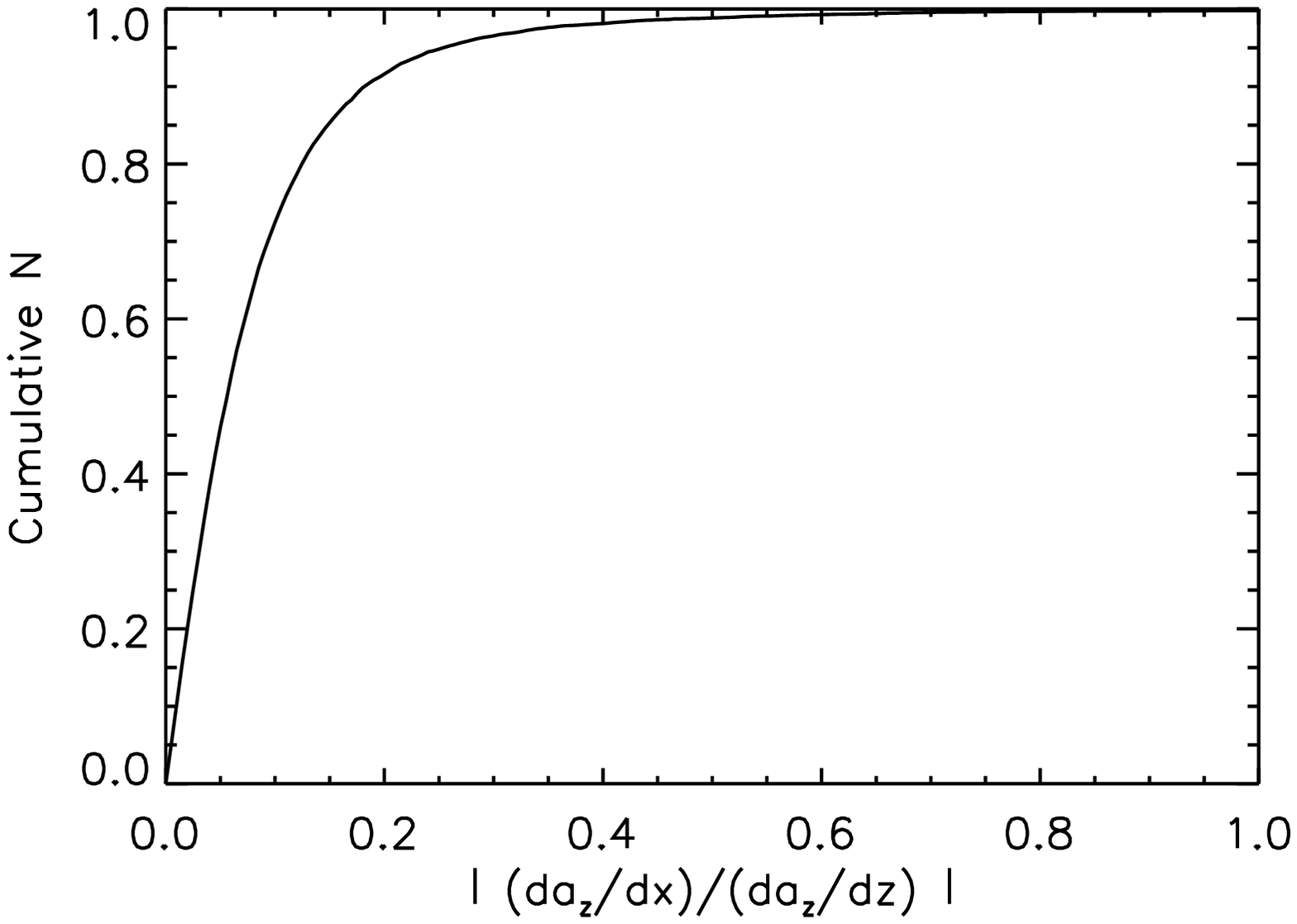}
\caption{Cumulative histogram of $\left |\frac{da_z}{dx}/\frac{da_z}{dz}\right |$ for the local tidal data in our simulated solar analogs.}\label{fig:tidetest1}
\end{figure}

Our first exercise is shown in Figure \ref{fig:tidetest1}.  Here we show a cumulative histogram of the ratio of the vertical gradient of the vertical acceleration ($\frac{da_z}{dz}$) to the  ${\bf \hat{x}}$-gradient ($\frac{da_z}{dx}$) (the ratio for the ${\bf \hat{y}}$-gradient has a nearly identical shape).  The data shown here is for all data outputs for every solar analog.  Because the analytical tides used in most Oort Cloud studies are dominated by the tide of the galactic disk, the vertical acceleration should not vary over small distances in the ${\bf \hat{x}}$ and ${\bf \hat{y}}$ directions.  As can be seen in Figure \ref{fig:tidetest1}, for over 90\% of our tidal data, the vertical acceleration varies by a factor of at least 5 slower for shifts in the ${\bf \hat{x}}$-direction than in the ${\bf \hat{z}}$.  Thus, it seems that our numerical tidal data generally behaves similarly to analytical approximations in this respect.

\begin{figure}[tbp]
\centering
\includegraphics[scale=.87]{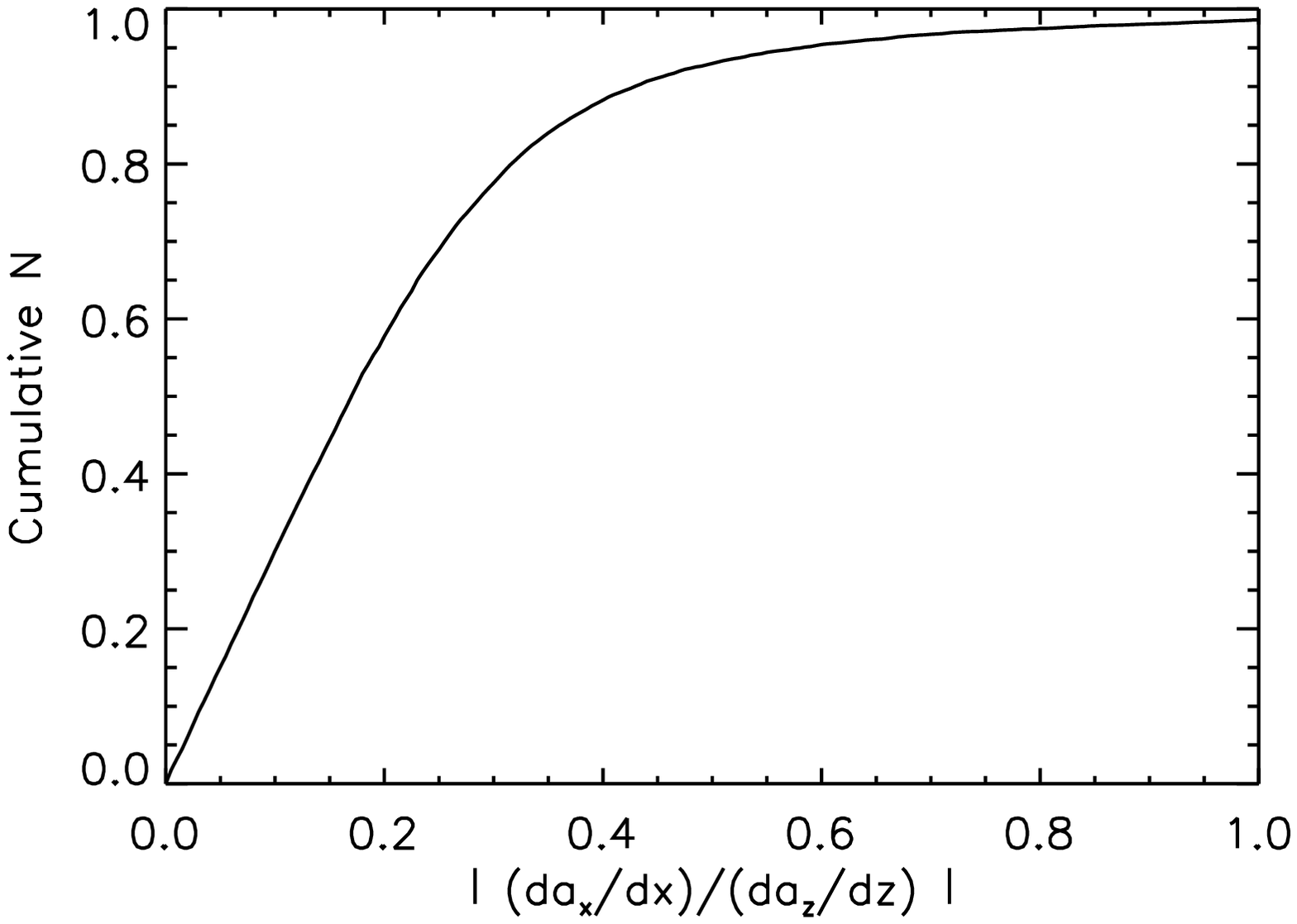}
\caption{Cumulative histogram of $\left |\frac{da_x}{dx}/\frac{da_z}{dz}\right |$ for the local tidal data in our simulated solar analogs.}\label{fig:tidetest2}
\end{figure}

The next examination of our tidal field calculations is shown in Figure \ref{fig:tidetest2}.  In this figure, we show another cumulative histogram, this one displaying the ratio of the ${\bf \hat{x}}$ gradient of the ${\bf \hat{x}}$ acceleration ($\frac{da_x}{dx}$) to the vertical gradient of the vertical acceleration ($\frac{da_z}{dz}$).  For analytical approximations of the Galactic tide near the Sun, the vertical tide dominates over tangential and radial components by a factor of 5-10 \citep{lev01}.  Once again, we see that our tidal data generally shows the same trends.  For 80\% of our tidal data, the vertical tidal component is at least a factor of $\sim$3 times stronger than ${\bf \hat{x}}$ component.  Furthermore, most of the tidal data where this ratio is large occurs when the solar analog in question has an excursion of several hundred pc away from the disk midplane, where the local disk density (and therefore $\frac{da_z}{dz}$) is lower.  As in Figure \ref{fig:tidetest1}, a tidal histogram where the ${\bf \hat{y}}$ tidal component is substituted for the ${\bf \hat{x}}$ component looks nearly identical to the one shown.  

\begin{figure}[tbp]
\centering
\includegraphics[scale=.87]{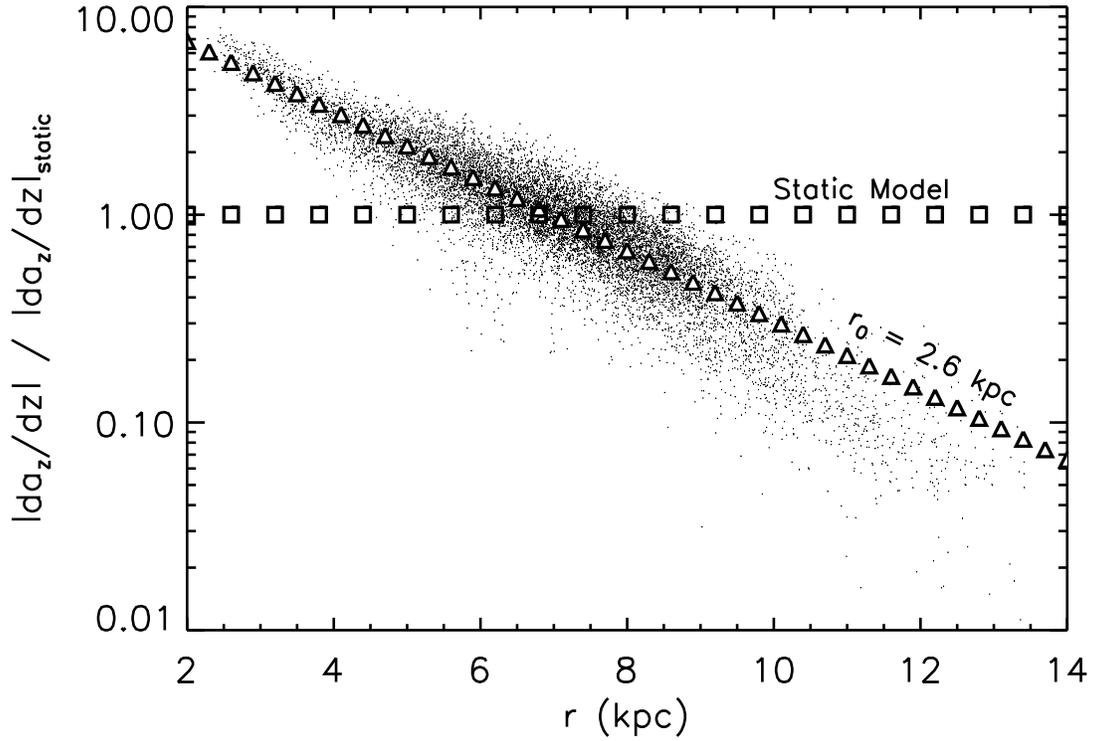}
\caption{Plot of $\left |\frac{da_z}{dz}\right |$/$\left |\frac{da_z}{dz}\right |_{\rm static}$ vs. galactocentric distance for the local tidal data in our simulated solar analogs (small points).  The line of square data points marks the value of $\frac{da_z}{dz}$ for an analytical tidal model \citep{lev01}, and the triangular points mark the best fit disk scale length for our tidal data.  The best fit yields a scale length of 2.6 kpc.}\label{fig:tidetest3}
\end{figure}

In the last assessment of our tidal field measurements we compare the local vertical gradient of the vertical acceleration ($\frac{da_z}{dz}$) with the distance from the Galactic center where the tidal measurement was taken.  In a galactic tidal field that is dominated by the disk tidal term, the vertical tide will be directly proportional to the local disk density.  As one moves away from the Galactic center, the disk density falls off exponentially, so we expect that the vertical tidal will show a similar trend.  Indeed, Figure \ref{fig:tidetest3} confirms that the local vertical tide we measure is strongly dependent on the distance from the Galactic center.  (Note that a plot of $\frac{da_z}{dz}$ vs $|z|$ does not show such a dependence due to the lesser significance of vertical oscillations explained in the previous section.)  For reference, we also mark the vertical tide given by a static analytical approximation used in several recent Oort Cloud works \citep{bras06,lev01,wietre99}.  In addition, we fit an inferred disk scale length to our tidal data.  We find a best fit scale length of 2.6 kpc, which falls well within the range of scale lengths we measure in Table \ref{tab:gal}.  We also see in Figure \ref{fig:tidetest3} that the maximum measured $\frac{da_z}{dz}$ values are near the analytical tidal model at 8 kpc (the approximate solar radius).  The lower measured values at 8 kpc correspond to excursions from the midplane or earlier times in the simulated galaxy's evolution when disk density is lower.  In light of the previous three figures, we believe that our numerical tidal calculation technique is suitable for exploring the effects of radial stellar migration on Oort Cloud dynamics.

\subsection{Oort Cloud Simulations}

We select 31 stars in the last timestep of our galaxy simulation that finish with galactic positions, velocities, and ages similar to the Sun.  These stars display a wide variety of dynamical histories in the galaxy, and we use direct numerical simulations to model the orbital evolution of test particles around these stars.  In addition, we perform three ``control'' simulations to mimic the typical assumption for the the solar neighborhood - the tidal field and stellar population at $r=8$ kpc in the last timestep of our simulated galaxy is used for the entire time.  To perform these simulations, we use the numerical algorithm SCATR \citep{kaib11a}.  This code integrates massless test particles with an adaptive timestep routine whose accuracy is enhanced with a symplectic corrector \citep{wis06}, making it particularly adept at modeling problems with a wide-range of dynamical timescales, such as the scattering and torquing of small bodies in the distant solar system.  Particles within 300 AU of the solar system barycenter are integrated with a timestep of 200 days, and particles beyond this distance are integrated with 9,000-day timesteps, greatly increasing the computing efficiency of distant orbits.

In our numerical simulations, we represent small bodies with massless test particles.  In each simulation, we begin with 2,000 test particles orbiting between 4 and 40 AU.  Their initial eccentricities are randomly distributed between 0 and 0.01, while the cosines of their initial inclinations are randomly distributed between 0 and 0.02.  These particles are then evolved under the gravity of the Sun and the four giant planets on their present orbits.  In addition to the Sun and giant planets, we also include gravitational perturbations from the galactic tide and passing field stars.  The galactic tide we employ is dependent on six independent terms: the ${\bf \hat{x}}$, ${\bf \hat{y}}$ and ${\bf \hat{z}}$ derivatives of each component of the galactic acceleration ($a_x$, $a_y$, and $a_z$) (not nine independent terms, since $\frac{da_{x}}{dy} = \frac{da_{y}}{dx}$, etc.).  These are recorded in our galaxy simulation every 10 Myrs and are updated accordingly in the Oort Cloud formation simulations.  In the time between tidal updates, the tidal terms are linearly interpolated between the last update and the next.  

In addition to tidal information, our galaxy simulation code also records the local stellar density and velocity dispersion near each of our solar analogs.  These quantities are determined from the 64 nearest stellar neighbors of the solar analog in question.  The recorded stellar velocity dispersion values range from 21 to 165 km/s, while local stellar densities are measured to be between 5 x 10$^{-3}$ and 1.4 M$_{\sun}$/pc$^3$, although the vast majority (99\%) are less than 0.5 M$_{\sun}$/pc$^3$.  The highest stellar densities and dispersions occur when solar analogs are closest to the Galactic center.  In our Oort Cloud simulations, random stellar encounters are generated by starting stars at random orientations 1 pc from the Sun.  These stars are then given random velocities using the encounter recipe of \citet{rick08} where stars are assigned different encounter velocities using the mass-velocity relation from \citet{garc01}.  It should be noted that this velocity relation is continuously scaled up or down so that the mean dispersion will match the local dispersion measured in our galaxy simulation, which is updated every 10 Myrs.  Lastly, the stellar encounter frequency is set by the locally measured dispersion and density of the galaxy simulation and is also updated every 10 Myrs.

Because the vast majority of particles scattered by the giant planets are ultimately ejected from the solar system, only a small percentage of our original 2,000 test particles will be bound to the Sun and torqued out of the planetary region after 4 Gyrs.  The signatures of different solar dynamical histories may be difficult to characterize with such low particle numbers.  For this reason, when our test particles pass through critical stages of scattering and torquing they are cloned with a routine similar to previous works \cite[e.g., ][]{levdun97, lev01}.  The first time particles are cloned is when they are being scattered by the giant planets to larger semimajor axes.  When a particle attains $a >$ 100 AU, it is cloned 10 times by shifting each of its cartesian coordinates between $\pm$1 x 10$^{-7}$ AU.  Then, if a particle's perihelion is torqued out of the planetary region ($q > 45$ AU) it is again cloned 10 times.  Through this process we generate 100 times as many particles with perihelia beyond the planets as an uncloned simulation would, resulting in much smoother distributions of orbital elements.  

\subsection{Stellar Encounter Monte Carlo Simulations}
\label{sec:monte}

In the present work, we also include a third set of Monte Carlo simulations modeling stellar perturbations.  These simulations are motivated from results of our planetary N-body simulations.  In the analysis of our N-body simulations, we pay special attention to the parameter $a_{sed}$, which is defined as follows:  Considering only bodies with Sedna-like perihelia (60 AU $<q<$ 100 AU), $a_{sed}$ is the semimajor axis inside which only 10\% of orbits are found.  From our N-body simulations, we observed that stellar encounters are responsible for setting the value of this parameter (see next section).  Because stellar passages are a stochastic process, it is impossible to determine the probability of Sedna production for a given solar orbital history with just one N-body simulation.  However, the computing requirements to perform many N-body simulations for each solar orbital history would be prohibitive.  Thus, we have run many simple Monte Carlo simulations of stellar encounters to complement our N-body simulations.

Fortunately, the total impulse delivered by passing stars is dominated by the few strongest encounters \citep{rick76}, so we only need to consider a handful of stellar encounters over the course of 4.5 Gyrs.  To measure the variance of the very inner edge of the Oort Cloud ($a_{sed}$) we generate 100 different sets of stellar encounters for each of our 31 solar orbital histories we modeled with N-body simulations.  In each set of stellar encounters, we ignore all encounters except the 5 closest encounters and the 5 encounters that deliver the largest velocity kick to the Sun.  We note that the median encounter timescale ($b/v$) of our set of powerful encounters is 104 yrs, whereas the orbital period of Sedna's orbit is much larger at $\sim$11000 yrs.  Thus, we can use the impulse approximation to model the perturbations of most of these stellar encounters on Oort Cloud bodies.  Occasionally included in our set of encounters are massive stars encountering the solar system at quite large distances and therefore much larger encounter timescales, which invalidate the impulse approximation.  For these encounters, if the encounter timescale is longer than 1000 yrs, we discard the encounter.  Therefore, although we know from observing our N-body simulations that it is usually the closest stellar encounters that alter $a_{sed}$, the results from our Monte Carlo simulations should be considered a minimum estimate of the effects of field stars.  

Once we have isolated the strongest encounters for each 4.5-Gyr history of encounters, we calculate their additive impulse on a generic scattered disk and Oort Cloud.  The generic scattered disk/Oort Cloud configuration we use is a $t=1$ Gyr snapshot from one of our control simulations.  After the impulse approximation is applied to our particles, we calculate the new modified orbital elements in our scattered disk and Oort Cloud.  As with the N-body simulations, we now consider only particles with 60 AU $< q <$ 100 AU and determine the value of $a_{sed}$.  Thus, with a small number of calculations for each stellar encounter set, we now have 100 different $a_{sed}$ values for each solar orbital history.  

Compared to our N-body simulations, these simulations are of course very simple, as they do not include the effects of the planets or Galactic tide.  However, this is an acceptable approximation for bodies with Sedna-like semimajor axes for the following reason.  If a low-$a$ orbit happens to be torqued out of the planetary region by one of the most powerful stellar encounters, it will no longer be influenced by planetary perturbations.  Additionally, the smallest semimajor axes affected by powerful stellar encounters will reside close enough to the Sun that the effect of the Milky Way's tide will also be negligible (again, see next section).  Thus, such an orbit will not evolve further until a stellar encounter of similar strength occurs, which we do account for.  

Perhaps the weakest aspect of these simulations is our choice of a generic scattered disk and Oort Cloud.  The reason for this is that although the dynamics of Sedna-like bodies are governed solely by the few strongest stellar encounters, this is not true for more distant parts of the Oort Cloud.  The Milky Way's tide sculpts the structure of most of the Oort Cloud, and the strength of this tide will vary significantly for different solar orbital histories.  Yet, all of our Monte Carlo stellar encounter simulations have the same Oort Cloud.  Essentially, we are accurately modeling the torquing of bodies out of the scattered disk (since the structure of the scattered disk will be nearly independent of external perturbations), and then we are tacking on an artificial Oort Cloud.  However, including some type of Oort Cloud is necessary since distant Oort Cloud orbits also occupy the 60 AU $< q <$ 100 AU region, and their absence would dramatically alter the location of $a_{sed}$.  

\section{Simulation Results}
\label{sec:galres}

For reference, we display a summary of our simulations in Table \ref{tab:runs}.  Included are the main orbital characteristics of each solar analog as well as the main characteristics of the resulting Oort Cloud.  What follows is a more detailed presentation of our simulation results that is divided into two subsections.  The first focuses on the production of Sedna-like orbits, while the second examines the entire Oort Cloud formed around each solar analog.

\subsection{Production of Sedna Analogs}

\begin{figure}[tbp]
\centering
\includegraphics[scale=.9]{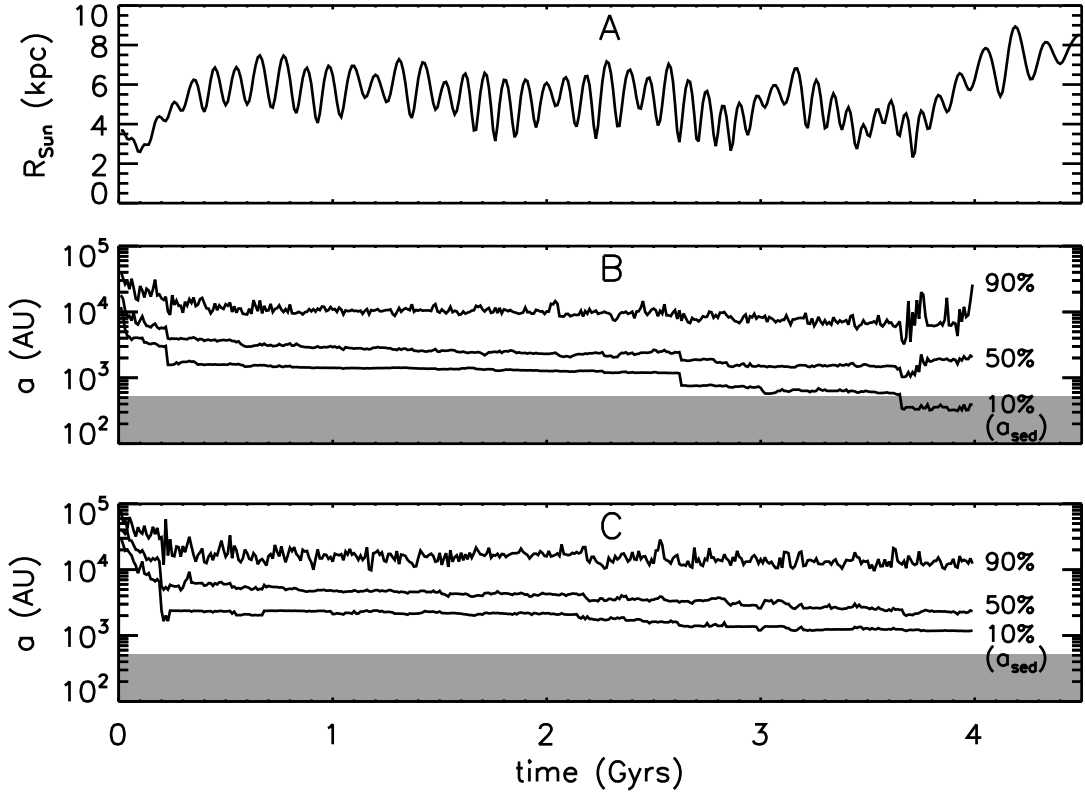}
\caption{{\it\bf a:} Plot of galactocentric distance vs. time for a typical solar analog. {\it\bf b:} Plot of the $a$ distribution for orbits with 60 AU $< q <$ 100 AU vs. time for the above solar analog.  The solid lines mark the semimajor axis inside which 10\%, 50\%, and 90\% of orbits are found. (Note that the 10\% curve defines $a_{sed}$.)  The shaded region marks semimajor axes inside Sedna's.  {\it\bf c:} Plot of the same orbital distribution as in {\it\bf b}, but with a fixed galactocentric distance of 8 kpc assumed.}\label{fig:scifig1}
\end{figure}

The potential effects of radial galactic migration on the outer solar system are illustrated in Figure \ref{fig:scifig1}.  Figure \ref{fig:scifig1}a shows the evolution of one particular solar analog's galactic orbit with time.  Although it eventually attains a solar-like galactic orbit, it spends its first 4 Gyrs significantly closer to the Galactic center where the average local density is about a factor of 4 higher.  Consequently, the external perturbations from the local environment are more powerful for a star with this orbital history.  In Figure \ref{fig:scifig1}b, we study the evolution of test particles orbiting this star.  Specifically, we follow the semimajor axis distribution of bodies that are torqued just beyond the planets (60 AU $< q<$ 100 AU).  In particular, we pay attention to $a_{sed}$, the semimajor axis inside which only 10\% of these particles are found.  Because this parameter is near the minimum semimajor axis torqued by external perturbations, $a_{sed}$ effectively marks the inner edge of the Oort Cloud. Figure \ref{fig:scifig1}b indicates that after 4 Gyrs, objects with semimajor axes even smaller than Sedna's are torqued beyond the planets.  In contrast, a simulation that assumes the Sun's galactocentric distance has remained fixed (Figure \ref{fig:scifig1}c) does not yield Sedna analogs.  Thus, for some solar migration cases, the Oort Cloud's very inner edge can be pushed to encompass Sedna's semimajor axis, making it one of the innermost members of the ``normal" Oort Cloud.  Of the 31 solar analogs for which we ran full simulations, 7 had $a_{sed} < 600$ AU after 4 Gyrs of evolution.

When examining the evolution of $a_{sed}$ in Figure \ref{fig:scifig1}b, one notices that $a_{sed}$ does not change smoothly.  Instead it is nearly constant for hundreds of Myrs before decreasing abruptly in very short periods.  If the Galactic tide were controlling the position of $a_{sed}$ we would expect to see a much more gradual evolution in $a_{sed}$ that reflects the steady nature of this perturbation.  Even for a migrating star, significant changes in radial position occur over $\sim$100 Myrs or longer, so the decreases in $a_{sed}$ should occur over these timescales as well if the tide were setting the Oort Cloud's inner edge.  

The only alternative to explain the evolution of $a_{sed}$ is that stellar perturbations are setting the position of $a_{sed}$.  This is quite surprising, as the Galactic tide has been shown to be more powerful than stellar perturbations when averaged over long timescales \citep{mormul86,heistre86}.  However, recent modeling of Oort Cloud dynamics suggests that the importance of stellar perturbations has been underestimated in many other recent studies \citep{rick08}.  Furthermore, stellar encounters are a stochastic process whose net effects are dominated by the strongest few encounters \citep{rick76}.  As a result, it is possible that the net impulse delivered by all stellar perturbations is concentrated into a few small time windows during which the most powerful stellar passages occur.  During these small windows, we would expect stellar perturbations to temporarily dominate over tidal perturbations, torquing bodies out of the planetary region on semimajor axes that are smaller than those expected due to just the Galactic tide.

To test whether this scenario is accurate, we perform a simple experiment.  First, we randomly generate a set of 100 scattered disk orbits around the solar analog studied in Figure \ref{fig:scifig1}b. The semimajor axes and perihelia are all fixed at 1,000 AU and 40 AU respectively, while inclinations are distributed between 0 and 10 degrees.  Because bodies with such eccentric orbits spend most of their time near aphelion, we fix the mean anomalies to $\pi$.  All other orbital elements are chosen randomly from a uniform distribution.  After these initial orbits are generated, we next use the impulse approximation to simultaneously apply all of the velocity impulses due to every stellar encounter within a 100-Myr window of our 4-Gyr N-body simulation.   This is done for every 100-Myr interval of our simulation, and for each interval, the mean absolute perihelion shift of our suite of orbits is recorded.  Similarly, we also calculate the mean absolute perihelion shift due to tidal forces for each interval by integrating the tidal acceleration experienced during each 100-Myr interval.  

\begin{figure}[tbp]
\centering
\includegraphics[scale=0.9]{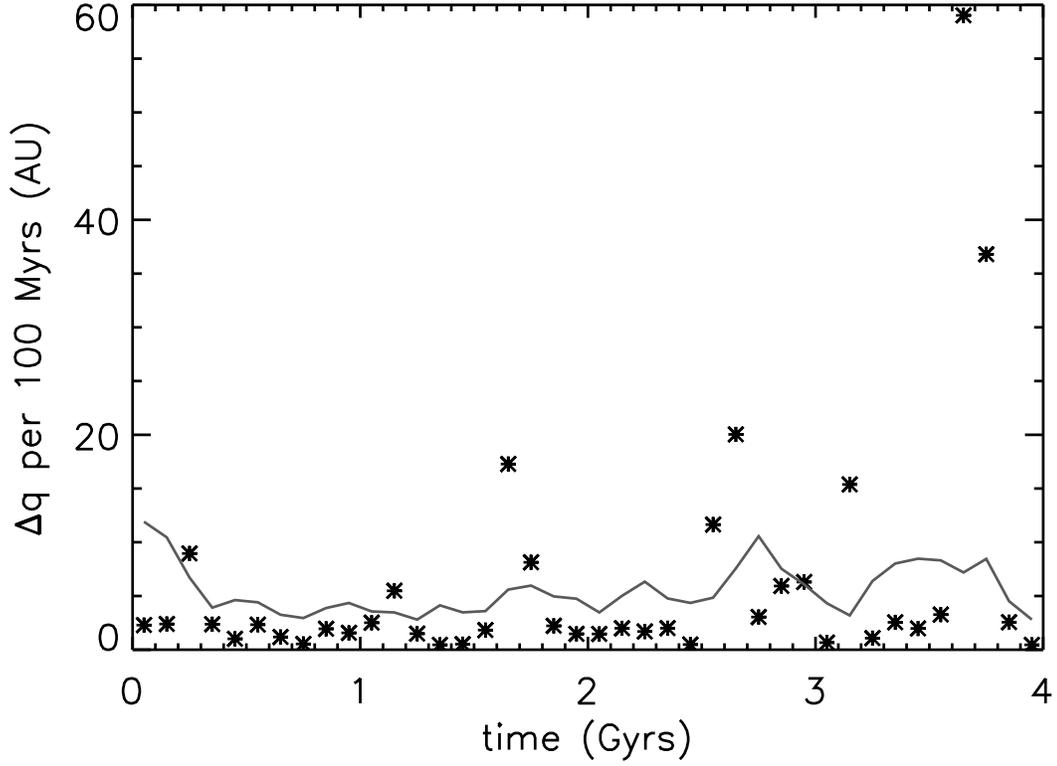}
\caption{Mean absolute perihelion change vs. time for 100 particles with $a=1000$ AU and $q=40$ AU.  Data points mark the perihelion change due to stellar perturbations, while the solid line marks the perihelion shift due to the Galactic tide.  This is the same set of perturbations experienced by the solar analog plotted in Figs. \ref{fig:scifig1}a and \ref{fig:scifig1}b.}\label{fig:pertcomp}
\end{figure}

The results of this perturbation study are shown in Figure \ref{fig:pertcomp} where we have plotted the mean perihelion shift due to stellar encounters vs. time as well as the mean perihelion shift due to the Milky Way tide vs. time.  This plot shows that, indeed, roughly 75\% of the time, tidal perturbations are more powerful than stellar perturbations.  However, there are also windows of time when a single very powerful stellar encounter greatly enhances the net effect of stars during a particular 100-Myr window.  During these windows, the perturbations of stars can be nearly 10 times as powerful as the tides.  Reexamining Figure \ref{fig:scifig1}b, we see that $a_{sed}$ experiences its sharpest decreases at $t=0.22$, 2.63, and 3.65 Gyrs.  Figure \ref{fig:pertcomp} shows that all of these decreases occur during intervals dominated by stellar encounters.  This verifies that stellar encounters set the location of the Oort Cloud's inner edge.

\begin{figure}[tbp]
\centering
\includegraphics[scale=0.9]{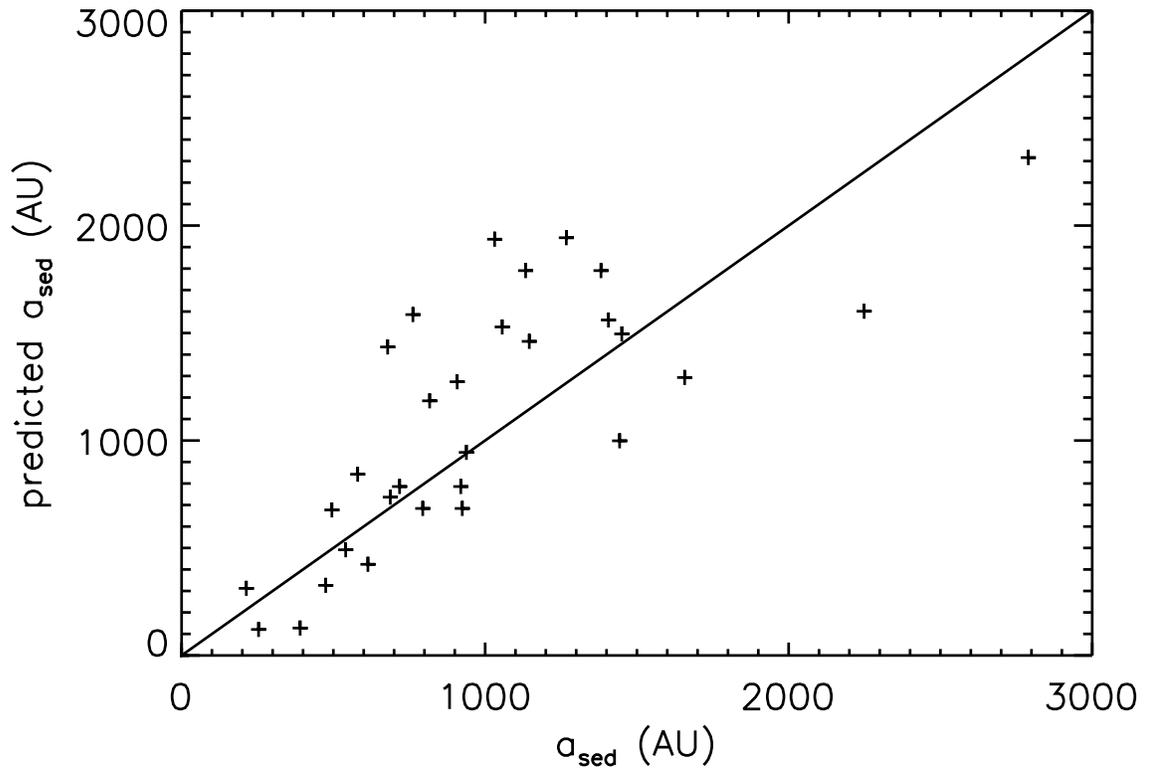}
\caption{Plot comparing the $a_{sed}$ predicted by our Monte Carlo stellar encounter simulations vs. the $a_{sed}$ measured in our N-body simulations using the same stellar encounter sets.  The solid line marks a 1:1 agreement.}\label{fig:imptest}
\end{figure}

It appears only a handful of isolated encounters set the location of the Oort Cloud's inner edge.  Due to these events' stochastic nature, the inner edge's location will vary substantially from one simulation to another, even in the same stellar environment.  Thus, there is always a spread of possible inner edge locations for a given stellar density, and it is impossible to measure this spread for a given solar analog using just one N-body simulation.  Instead, we can use our Monte Carlo simulations (see Section \ref{sec:monte}) to help predict the range of $a_{sed}$ for the stellar environments encountered by each solar analog.  First, however, we must assess how well these additional Monte Carlo stellar encounter simulations predict the position of $a_{sed}$.  To do this, we perform stellar encounter simulations that use the same exact stellar encounter sets from our N-body simulations.  Then we compare the $a_{sed}$ values predicted from the stellar encounter simulations with those measured in our full N-body simulations.  The results of this test are shown in Figure \ref{fig:imptest}.  We see that although there is scatter, the 1:1 trend demonstrates our Monte Carlo simulations are capturing the dominant process responsible for setting the inner Oort Cloud edge.  (The scatter is most likely due to our use of a generic scattered disk/Oort Cloud configuration described in Section \ref{sec:monte}.)

\begin{figure}[tbp]
\centering
\includegraphics[scale=.9]{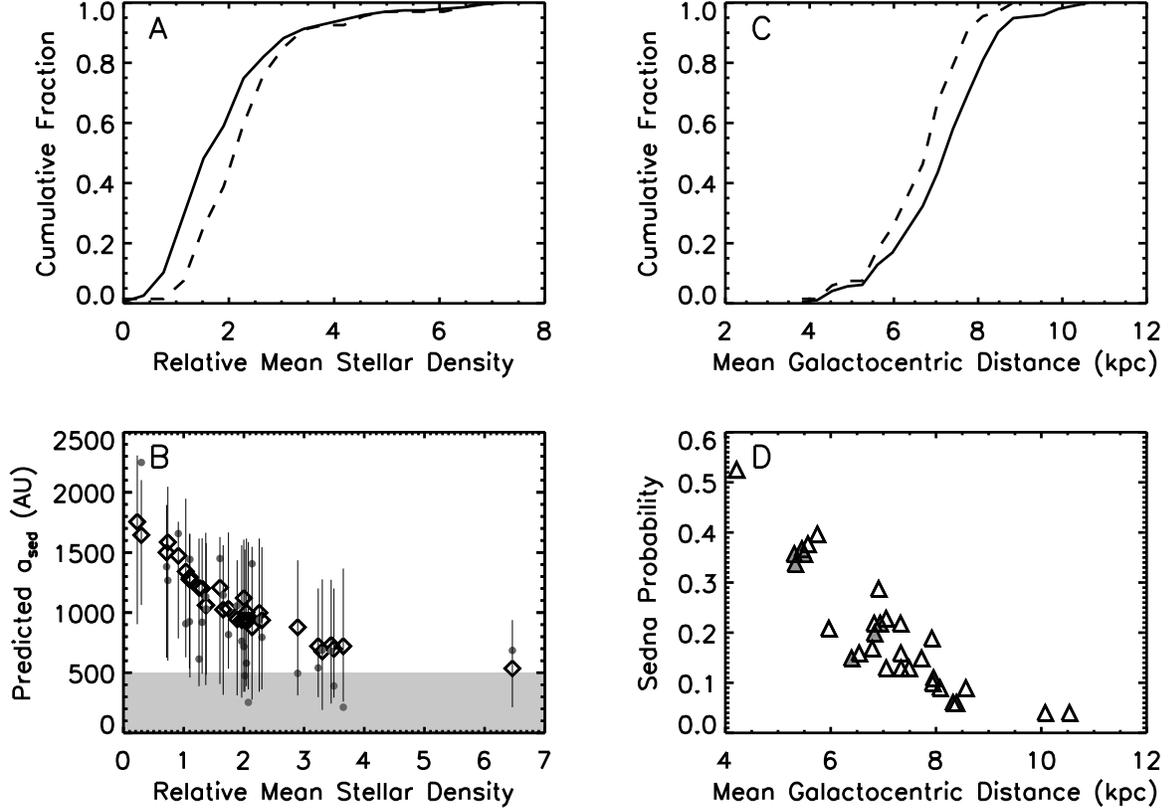}
\caption{{\it\bf a:} Cumulative distribution function (CDF) for the mean local stellar density encountered by our solar analogs relative to the present local stellar density. The CDFs for all analogs ({\it solid}) and near-solar metallicity analogs ({\it dashed}) are shown.  {\it\bf b:} Predicted a$_{10}$ vs. mean relative local stellar density encountered by solar analogs.  Open diamonds mark the median value from our monte carlo impulse approximation simulations, while the error bars mark the range of the middle 80\% of $a_{sed}$ values generated for each analog. Solid dots mark the $a_{sed}$ values from full N-body simulations.  The shade region marks semimajor axes less than Sedna's.  {\it\bf c:} CDF for the mean galactocentric distances of all solar analogs ({\it solid}) and solar metallicity analogs ({\it dashed}).  {\it\bf d:} The probability that Sedna analogs are produced ($a_{sed} < 600 $ AU) vs. mean galactocentric distance for our solar analogs.  Shaded data points correspond to N-body simulations solar that actually attained $a_{sed} < 600$ AU.  }\label{fig:scifig2}
\end{figure}

We perform 100 Monte Carlo simulations of random stellar encounters to complement each of our N-body simulations, and classify a simulation as "Sedna-producing" if $a_{sed}$ is pushed inside 600 AU.  If we assume the current solar neighborhood has been unchanged for 4.5 Gyrs, Monte Carlo simulations for our control cases indicate that there is an $\sim8$\% chance that Sedna analogs can result from field star torques.  Even this is a surprisingly high rate, considering most previous Sedna works have discounted such a scenario.  However, when we examine the mean stellar densities encountered by all of our solar analogs in Figure \ref{fig:scifig2}a we see that the vast majority of solar analogs spend most of their time in regions with even higher stellar densities than our control case.  Consequently, this spread of possible inner edges is shifted to lower semimajor axes for most solar orbital histories.  As Figure \ref{fig:scifig2}b shows, the median location of the Oort Cloud's inner edge gets steadily closer to Sedna's orbit for solar histories with higher mean stellar densities, with Sedna becoming the median case for our most extreme analog.  As can be seen in Figure \ref{fig:scifig2}d, this also translates into a tight correlation between the probability of Sedna production and the Sun's average distance from the Galactic center (since the local stellar density is strongly dependent on this distance).  Depending on the Sun's orbital history, this probability can range between 2 and $\sim$50\%!  Because our simulated galaxy contains a chemical enrichment routine \citep{stin06}, we can also use the metallicity of our solar analogs to gauge which orbital histories are most likely.  We see in Figures \ref{fig:scifig2}a and \ref{fig:scifig2}c that selecting only analogues with near solar metallicity ($\pm0.1$ dex) biases our sample toward histories with higher stellar densities and smaller galactocentric distances.  (This is due to the fact that higher metallicity stars are more likely to form from more metal-rich gas nearer to the Galactic center.)

Our result that $a_{sed}$ can be pushed inside a few thousand AU is in contrast to many previous Oort Cloud studies.  These differences can mostly be explained by the variable nature of our galactic environment.  The exception to this is the work of \citet{bras10}, which modeled the formation of the Oort Cloud at various distances from the Galactic center.  In particular, they ran three simulations at 2, 4, and 6 kpc from the Galactic center, and found that only bodies with semimajor axes beyond 1000 -- 2000 AU could be torqued beyond the planets.  There are several possible explanations for this result.  The first is that it may just be a matter of small number statistics.  Less than 30\% of our simulations generate Sedna analogs, while \citet{bras10} had a total of just three runs within the solar radius.  Second,  our models of the Milky Way are fundamentally different.  Our stellar and tidal perturbations are drawn from a numerical simulation, whereas theirs are based on an analytic approximation of the Milky Way potential.  

It is also important to note that we measure the inner edge of the Oort Cloud in a different manner than \citet{bras10}.  They require objects to attain $a>$ 1000 AU before considering them as Oort Cloud objects, so none of the particles they consider will have $a$ near Sedna.  Furthermore, we restrict ourselves to only orbits that have been weakly torqued beyond the planets (60 AU $< q <$ 100 AU) and measure the bottom tenth percentile in $a$.  Even for completely isotropized orbits, lower semimajor axes orbits are more likely to be found at low $q$ values, and our lowest semimajor axes orbits are far from isotropic, producing an even heavier low-$a$ bias at low $q$ values.  Consequently, relative to larger $a$ values, low-$a$ orbits densely populate the perihelion range of interest to us.  Meanwhile, \citet{bras10} looks at the lowest 5\% of all Oort Cloud semimajor axes spanning all possible perihelia.  This will include a much more isotropized sample of orbits at high-$q$, resulting in a more top-heavy $a$-distribution.  

\subsubsection{Delayed Sedna Formation}

Examining Figure \ref{fig:scifig1}b again, we see that $a_{sed}$  does not fall inside Sedna's orbit until very late in the solar system's history.  In our seven simulations that attain $a_{sed}<600$ AU, the mean time that $a_{sed}$ first drops below 600 AU is $t=2.2$ Gyrs, and only one simulation attains $a_{sed}<600$ AU within the first Gyr of evolution.  In light of this, Sedna's orbit could hardly be called primordial if it resulted from the effects of solar migration.  This is in contrast to perhaps the most well-studied Sedna production mechanism, that within a solar birth cluster.  In this scenario, Sedna must reach its orbit within the 5 Myrs it takes typical embedded clusters to disperse \citep{lala03}.  In this respect, our mechanism's delayed timing is compelling.  If the Sun was born into an embedded cluster environment that also enriched the solar nebula with supernova ejecta, then the time window to generate Sedna's orbit becomes uncomfortably small since supernovae seem to coincide with the dispersal of embedded clusters \citep{adams10}.  This time window is larger if the Sun was instead born into a longer-lived gravitationally bound cluster, but only a few percent of stars seem to originate in such environments \citep{lala03}.

Our model's predicted delay between solar system formation and Sedna production is also intriguing given that the formation of the modern scattered disk seems to have been delayed as well \citep{gom05,lev08}.  The planetary evolution model that most successfully explains the orbital architecture of the Kuiper Belt and scattered disk predicts this architecture was produced when a brief chaotic phase of the giant planets' orbital evolution scattered material throughout the entire solar system \citep{tsig05, lev08}.  This is thought to have occurred hundreds of Myrs after the planets first formed (and after the solar birth cluster dispersed) \citep{gom05}.  Consequently, any previous scattered disk from the original epoch of planet formation would have been quite depleted by this point and would have been blanketed with this newly scattered material.  Members of this new set of scattered bodies would be the most likely to wind up on Sedna-like orbits, since it typically takes over a Gyr for the proper set of powerful stellar encounters to occur in our simulations.

\subsubsection{Mass Requirements}

Our mechanism's delayed Sedna production also highlights a potential weakness, however.  In models producing Sedna within a birth cluster, Sedna-like orbits are generated very early, and any scattered disk that was present would be near its peak population.  During the proceeding 4 Gyrs, the scattered disk population would dynamically erode, whereas the Sedna population would remain intact since it is nearly unaffected by perturbing forces.  Therefore, the cluster model should predict relatively few Sedna-sized bodies in the modern scattered disk for each real Sedna.  On the other hand, our Sedna formation takes place after the scattered disk has already been eroded for at least hundreds of Myrs.  Consequently,  we require a much greater number of Sedna-sized bodies in the modern scattered disk.  To calculate this required number, we define a Sedna-like orbit as $a<$ 600 AU and 60 AU $<q<$ 100 AU, while we classify particles as scattered disk objects if they have 28 AU $<q<$ 40 AU and $e>$ 0.25.  Indeed, we find that for our 7 simulations with $a_{sed}<$ 600 AU, there should be between 4 and $\sim$90 Sedna-sized objects in the scattered disk for every similar-sized object in a Sedna-like orbit.  This is disturbing, as there are just two known scattered disk objects that are comparable in size to Sedna: Eris \citep{brown05} and 2007 OR$_{10}$ \citep{schwamb09}.

Although the predictions of our model are difficult to reconcile with current scattered disk observations, this may not be a flaw in our Sedna production mechanism, but perhaps our simplistic modeling of the scattered disk.  The formation of the real scattered disk most likely involved a chaotic evolution of the giant planets' orbits \citep{tsig05}.  Our simulations are somewhat crude in this respect, as they do not include such an evolution.  In addition, recent models of the early solar system suggest that the original disk of planetesimals was truncated at $\sim$30 AU \citep{gom05}.  By starting our particles on circular orbits between 4 and 40 AU, we have most likely overpopulated the classical Kuiper Belt.  Some of these particles will inevitably bolster the modern scattered disk population of our simulations.  (We note that 60\% of our Sedna-like objects had initial orbits inside 30 AU and not between 30 and 40 AU, where the original population may be artificially enhanced in our simulations.)

Unlike the scattered disk, we find that our mechanism's mass requirements on the original protoplanetary disk are more reasonable.  For our simulations that have $a_{sed}$ inside 600 AU, we find that for each Sedna that exists today we require between 800 and 30000 Sedna-sized bodies in the original protoplanetary disk.  If we assume a mass of 5 x 10$^{-4}$ M$_{\earth}$ for Sedna, this means that between $\sim$.5 and $\sim$17 M$_{\earth}$ of Sedna-sized bodies must have been in the original protoplanetary disk for each Sedna-like body today.  Such masses could be accounted for by current current models of solar system evolution \citep{gom05, morb07}.  

\subsubsection{Predicted Population Occupying the Sedna Region}

\begin{figure}[tbp]
\centering
\includegraphics[scale=0.9]{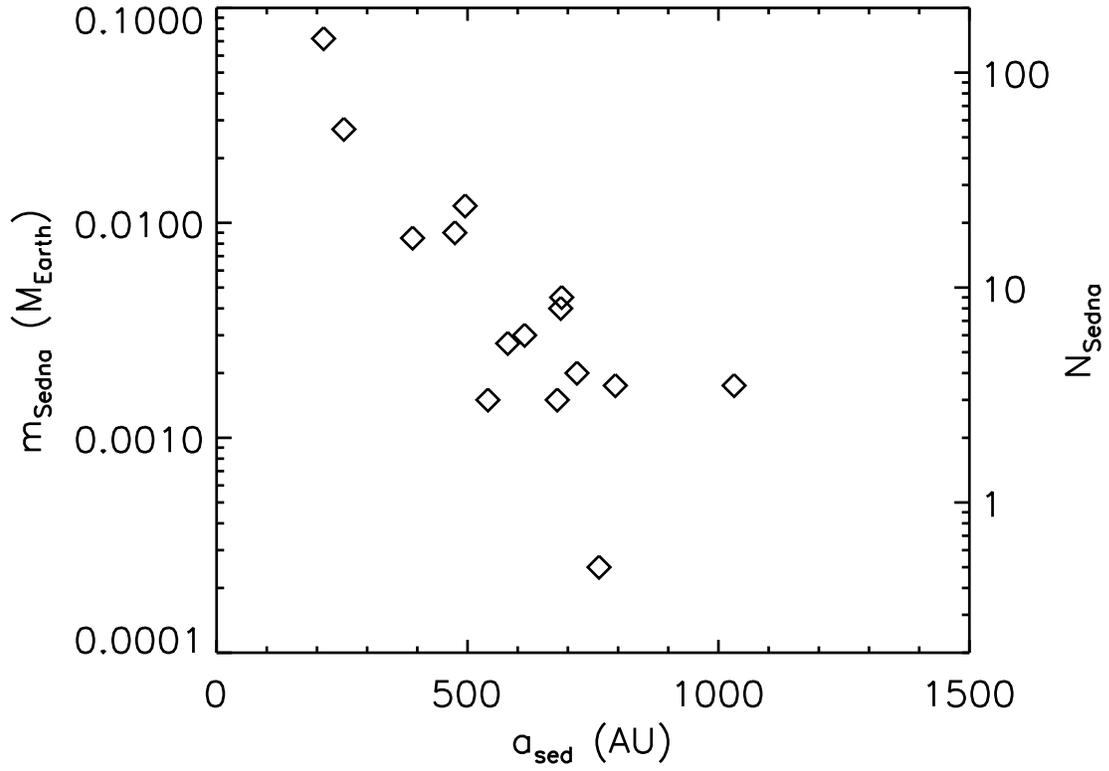}
\caption{Total mass of the population within the "Sedna region" vs. $a_{sed}$ for each one of our planetary N-body simulations.  We define the Sedna region as any orbit with $a<$ 600 and $q>$ 60 AU.  Assuming a mass of 5 x 10$^{-4}$ M$_{\earth}$ for Sedna, we convert this mass to "number of Sednas" on the righthand axis.}
\label{fig:sedmass}
\end{figure}

In the previous section, we examined the mass requirements placed on the scattered disk and protoplanetary disk by our models.  Conversely, if we assume a given value for the original mass of solids in the protoplanetary disk, we can also estimate the total mass of objects occupying the ``Sedna region" predicted by our simulations.  As a fiducial value for this exercise, we take our initial disk mass to be 50 M$_{\earth}$ since this is the disk mass favored by the most recent models explaining the orbital architecture of the giant planets and the formation of the modern scattered disk \citep{morb07}.  Assuming this disk value, the total mass in Sedna analogs found around each of our solar analogs is shown in Figure \ref{fig:sedmass}.  (To compute the masses in this figure, we count any object with $q>$ 60 AU and $a<$ 600 AU.)  We see in Figure \ref{fig:sedmass} that for our solar analogs with $a_{sed}\sim$400--600 AU, we would expect anywhere between a couple thousandths to $\sim$0.01 M$_{\earth}$ of material occupying Sedna-like orbits.  Once again assuming a mass of Sedna of 5 x 10$^{-4}$ M$_{\earth}$, this would be equivalent to $\sim$3 to $\sim$20 Sedna-like bodies.

\begin{figure}[tbp]
\centering
\includegraphics[scale=0.9]{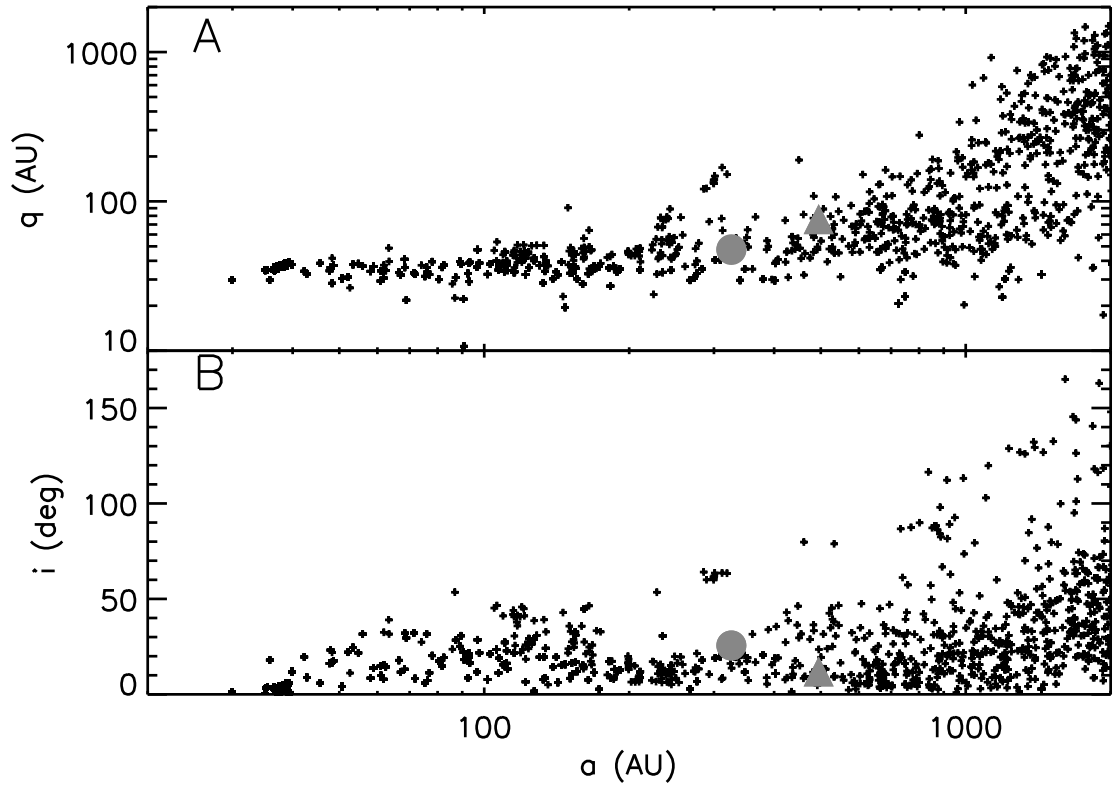}
\caption{{\it\bf a:} Plot of perihelion vs. semimajor axis for all test particles inside $a<$ 2000 AU in Run 4.  {\it\bf b:} Plot of inclination (relative to ecliptic) vs. semimajor axis for all test particles inside $a<$ 2000 AU in Run 4.  The orbit of Sedna is marked with a triangle in both plots, and the orbit of 2004 VN$_{112}$ is marked with a circle.}
\label{fig:simsamp}
\end{figure}

\begin{figure}[tbp]
\centering
\includegraphics[scale=.9]{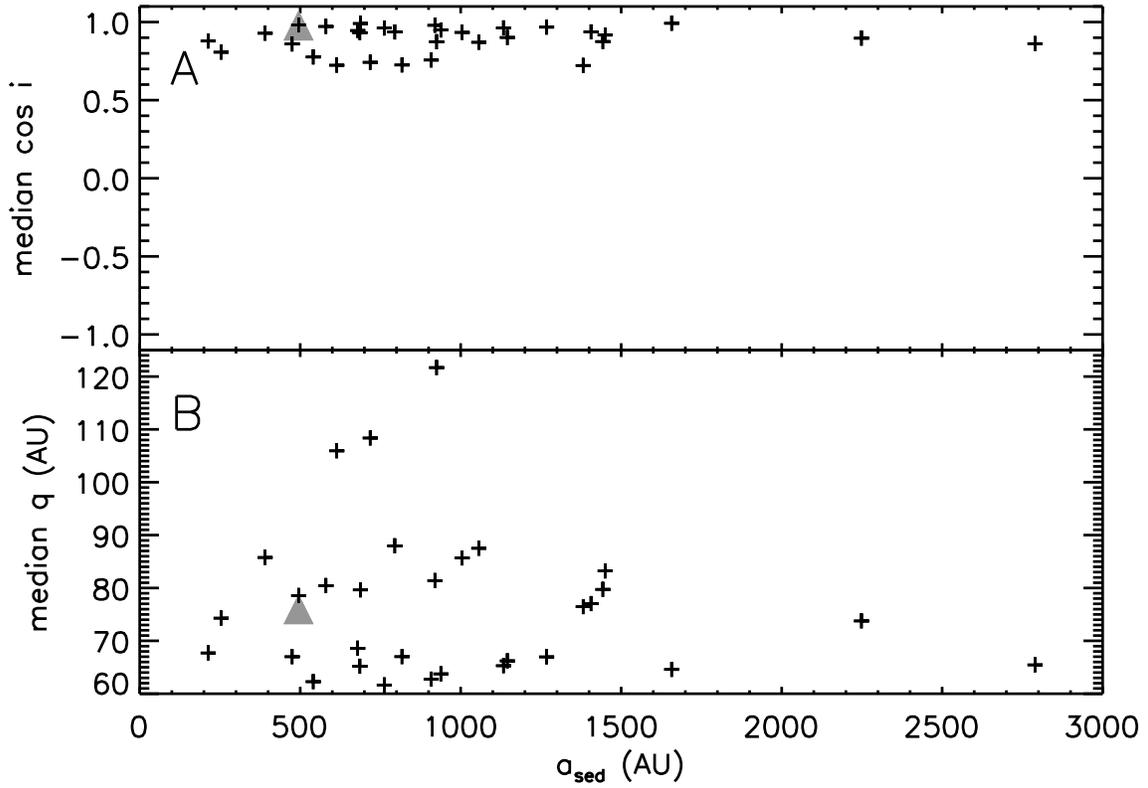}
\caption{{\it\bf a:} Plot of median $\cos{i}$ for particles with $a<a_{sed}$ and 60 AU$<q<$100 AU vs. $a_{sed}$ for each N-body simulation performed for a solar analog.  {\it\bf b:}  Plot of median perihelion for particles with $a<a_{sed}$ and $q>60$ AU vs. $a_{sed}$ for each N-body simulation. Sedna's orbit is marked with a triangle in each plot.}
\label{fig:scifig3}
\end{figure}

Although the total mass of Sedna analogs predicted by our models is small, it is important to realize that this population is highly anisotropic and concentrated in observable high eccentricity orbits.  In Figure \ref{fig:simsamp}, we display the orbits of all particles inside $a<$ 2000 AU for one of our nominal N-body simulations, Run 4 ($a_{sed}$ = 495 AU).  We can see that for particles with $a$ near Sedna, the orbital distribution is far from isotropic.  This contrasts with most models that produce Sedna within a star cluster environment.  For most cluster models the orbital distribution is nearly isotropic \citep{fernbrun00, bras06, kaibquinn08}.  This would imply that Sedna represents the first discovered member of a vast but mostly unobservable population whose eccentricities range fully between 0 and 1 and whose inclinations ($i$) are also nearly randomly distributed between 0 and 180 degrees.  On the other hand, our new mechanism is just strong enough to perturb the perihelion of Sedna beyond the reach of the planets.  Consequently, our sample of Sedna analogs has a very non-isotropic orbital distribution with inclinations strongly concentrated toward the ecliptic (see Figure \ref{fig:scifig3}a).  In addition, Figure \ref{fig:scifig3}b shows that for nearly all of our full N-body simulations, the majority of particles with $a<a_{sed}$ have perihelia inside 90 AU.  Thus, we predict no vast high-$q$ or high-$i$ population of Sedna analogs.  Although it can be misleading to infer a population's characteristics from a single known object, it is compelling that recent searches for Sedna analogs at high ecliptic latitudes have yielded no detections \citep{schwamb10}.  If upcoming synoptic surveys such as LSST and Pan-STARRS \cite[Ivezic {\it et al.} 2008 (arXiv/0805.2366);][]{kais02} detect a distribution of Sedna-like bodies similar to our predictions, this would suggest that our Sun has undergone substantial outward migration in the Milky Way.  On the other hand, the detection of retrograde Sedna-like orbits would be very difficult to explain with our mechanism.  

In addition to reproducing the orbit of Sedna well, Figure \ref{fig:simsamp} shows that this simulation also contains particles on orbits quite similar to 2004 VN$_{112}$.  With a perihelion of 47 AU and semimajor axis of $\sim$350 AU, 2004 VN$_{112}$ is very close to being dynamically isolated from the planetary region of the solar system, much like Sedna is.  However, its perihelion may be low enough that it could also possibly be an extreme example of perihelion-lifting due to Kozai interactions with Neptune \citep{gom05b}.  In our simulation, we see that there exist numerous test particles with perihelia well beyond $q\gtrsim50$ AU and with $a\sim350$ AU.  For these particles, their perihelia could only have been lifted by steller perturbations, and this raises the possibility that bodies like 2004 VN$_{112}$ may also be signatures of the Sun's radial migration.  Because it is less isolated from the planetary region, it is harder to quantify which of our runs generate 2004 VN$_{112}$ analogs and which do not.  Nevertheless, our N-body simulations with $a_{sed}\lesssim600$ AU seem to usually populate the 2004 VN$_{112}$ region, providing qualitative evidence that our mechanism can explain 2004 VN$_{112}$ as well.

Again due to its timing, our mechanism differs from Sedna production within a birth cluster in one other way.  Previous work indicates that aerodynamic drag due to the gaseous component of the solar nebula prevents bodies with radii below $\sim$10 km from reaching large $a$ before embedded clusters typically disperse \citep{bras07}, so this mechanism would imply an absence of km-sized bodies with Sedna-like orbits.  On the other hand, no size discrimination would occur if Sedna's orbit is due to the Sun's migration within the galaxy, as the most powerful field star passages almost certainly occurred after solar nebula dispersal.  Although direct observation of such small objects in distant orbits is currently impossible, such a prediction may be testable by stellar occultation experiments. 

\subsubsection{Simulated Galaxy vs. Milky Way}

Since the key ingredient of our Sedna formation mechanism is powerful stellar encounters, we must consider how much our powerful stellar encounters will differ between our simulated galaxy and the real Milky Way.  As noted in Table \ref{tab:gal}, our simulated galaxy has a higher velocity dispersion and stellar density at the solar radius than the present solar neighborhood.  To assess the effects of this, we perform 400 additional Monte Carlo stellar encounter simulations that assume a fixed stellar density of 0.04 M$_{\sun}$/pc$^{3}$ and a fixed dispersion of 42 km/s (values similar to the present solar neighborhood).  Of these simulations, $a_{sed}$ is pushed inside 600 AU 8\% of the time.  Remarkably, this is the same percentage as found from the Monte Carlo simulations for our control case that assumed a density of 0.056 M$_{\sun}$/pc$^{3}$ and a dispersion of 62 km/s.  The reason for this is that although the heightened dispersion and density of our simulated galaxy cause a greater number of encounters with the solar system as well as a lower expected minimum encounter distance, the increased stellar velocities decrease the encounter timescale and make each individual encounter weaker.  These two effects essentially cancel each other out.  Thus, to first order it appears that the most powerful encounters from our simulated galaxy will perturb the Solar System as much as the most powerful encounters from the real Milky Way.

\subsection{The Oort Cloud}
\subsubsection{Oort Cloud Trapping Efficiency}

We now turn our attention to the rest of the Oort Cloud formed around each of our solar analogs.  When analyzing the results of our planetary N-body simulations, we consider any body that attains a perihelion beyond 45 AU to be an Oort Cloud object.  After 4 Gyrs of evolution, we find that the characteristics of the Oort Cloud formed around each of our solar analogs is quite sensitive to that star's orbital migration history.  This migration causes significant changes in the population of passing stars as well as the Galactic tide.  As noted in \citet{rosk08b}, there is a general tendency for stars to radially migrate beyond their formation radius.  Consequently, many of our solar analogs experience more powerful tidal and stellar perturbations early in their history.  Because this increases the rate that perihelia can be torqued beyond the planetary region at a given semimajor axis, the number of bodies trapped in the Oort Cloud increases.  Typically, bodies scattered by Jupiter and Saturn experience such a rapid growth in semimajor axis that they are ejected before they can be trapped in the Oort Cloud.  With a shorter perihelion torquing timescale, however, a greater fraction of these bodies are trapped.  

\begin{figure}[tbp]
\centering
\includegraphics[scale=.87]{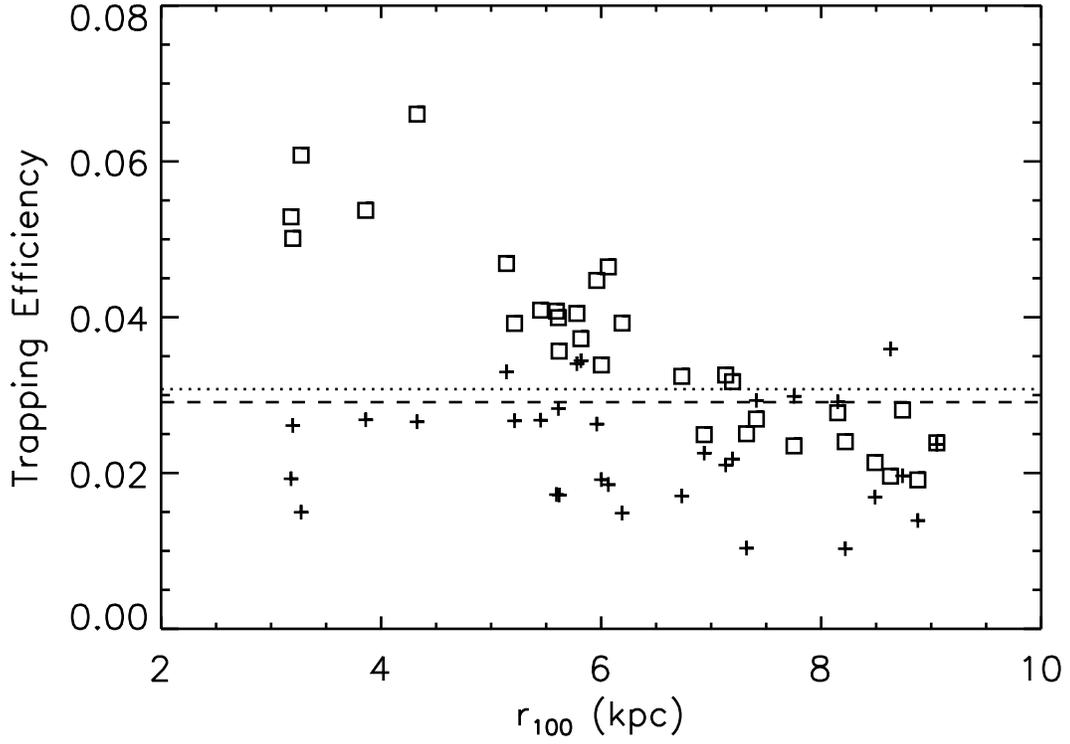}
\caption{Plots of Oort Cloud trapping efficiency vs. mean galactocentric distance during the first 100 Myrs of orbital evolution for each solar analog.  The square data points mark the trapping efficiencies measured after 100 Myrs, and the crosses mark the trapping efficiencies at 4 Gyrs.  In addition, we also show the 100-Myr and 4-Gyr trapping efficiencies in our control simulation with the dotted and dashed lines (respectively).}\label{fig:6}
\end{figure}

This effect is evident in Figure \ref{fig:6}, where we plot Oort Cloud trapping efficiency after 100 Myrs as a function of the mean galactocentric distance for this time period ($r_{100}$) for each solar analog.  We define the Oort Cloud trapping efficiency to be the fraction of original particles that are located in the Oort Cloud at a given time.  We see that solar analogs formed nearer to the Galactic center initially trap a higher fraction of material in the Oort Cloud.  Furthermore, the median value of $r_{100}$ (a proxy for stellar formation radius) is 6.1 kpc, indicating that most solar analogs formed inside the present solar orbit and experience this trapping enhancement.  However, this effect is short-lived.  The increased external perturbations will also strip the outermost orbits and enhance the rate that Oort Cloud bodies are re-injected into the planetary region where many are ejected by perturbations from Jupiter and Saturn.  Thus, once the initial population of planetesimals starting near Jupiter and Saturn is exhausted, the outer regions of the Oort Cloud are eroded by the enhanced tidal field faster than they can be replenished.  Only the Oort Cloud bodies in the innermost region of the cloud are protected from this effect.  Consequently, after 4 Gyrs we see in Figure \ref{fig:6} that all of the Oort Clouds we form have trapping efficiencies of between 1 and 3.5\% with no obvious dependence on the Sun's formation radius.  This supports the results of \citet{bras10} who showed that an Oort Cloud formed nearer to the Galactic center will have a higher initial trapping efficiency but will finish with one similar to that found near the solar radius.  

These results contrast with the works of \citet{bras06} and \citet{bras08a} who show that the tidal torques of an early embedded cluster environment can greatly enhance the Oort Cloud's trapping efficiency even after 4 Gyrs.  However, this is due to the fact that embedded clusters have very short lifetimes (1--5 Myrs).  As a result, the tidal forces of the cluster enhance the trapping efficiency of the entire Oort Cloud, but the cluster disperses before the supply of scattering bodies near Jupiter and Saturn is depleted.  Therefore, there is no time for significant cloud erosion to take place.  Simulations of Oort Cloud formation within longer-lived open clusters show that most of the trapping efficiency enhancement is lost by erosion due to the longer-lived perturbations \citep{kaibquinn08, fernbrun00}.  Similarly, because the dynamical timescale of stellar orbits in the Milky Way are hundreds of Myrs, any time period that the Sun is closer to the Galactic center will last far too long to increase the Oort Cloud's trapping efficiency.

Although most solar analogs form closer to the Galactic center and migrate outwards, this migration process is not always one way.  Before arriving at the Sun's current galactocentric distance, some solar analogs migrate inwards first, and the time at which they attain their minimum galactocentric distance ($t(r_{min})$) occurs later in their history.  In these cases, the stronger Galactic tide and stellar encounters experienced near the minimum galactocentric distance will have a different impact on the Oort Cloud than if $t(r_{min})$ occurred early in the star's history.  This is because the Oort Cloud is populated from the scattered disk, whose population decreases by a factor of $\sim$10 between $t=100$ Myrs and $t=2$ Gyrs.  As a result, a solar analog with a late $t(r_{min})$ will experience substantially less inner cloud enrichment than one with an early $t(r_{min})$.  On the other hand, because the population of the Oort Cloud changes more slowly than the scattered disk, both solar analogs will experience similar degrees of erosion in their outer clouds.  Hence, a late $t(r_{min})$ will be significantly more erosive than an early $t(r_{min})$.

\begin{figure}[tbp]
\centering
\includegraphics[scale=.87]{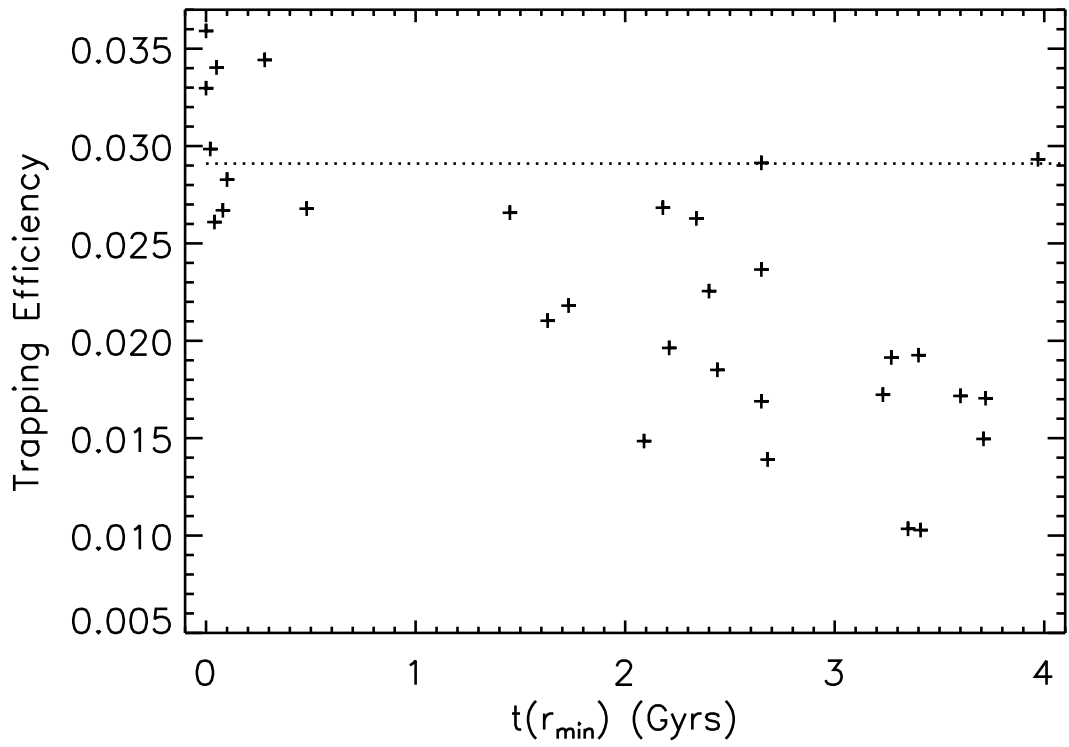}
\caption{Plot of Oort Cloud trapping efficiency vs. the time at which the simulated solar analog attains its minimum galactocentric distance. The trapping efficiency of our control simulation is shown by the dotted line.}\label{fig:tmin}
\end{figure}

This raises the question of whether Oort Cloud trapping efficiency can be tied to $t(r_{min})$.  Figure \ref{fig:6} demonstrates that although the trapping efficiency seemed to be independent of how close the Sun came to the Galactic center, there is still a spread of roughly a factor of 2 in the trapping efficiencies of our Oort Clouds after 4 Gyrs of evolution.  In Figure \ref{fig:tmin}, we plot the Oort Cloud trapping efficiency as a function of $t(r_{min})$.  We see that while there is still some scatter, there is a clear correlation between Oort Cloud trapping efficiency and $t(r_{min})$.  Solar analogs that attain their minimum galactocentric distance early in their history generally have a higher Oort Cloud trapping efficiency than those that have a late $t(r_{min})$.  The most exceptional outlier to this trend is the rightmost data point in this plot (Run 19).  In this case, the solar analog attains its minimum galactocentric distance only 30 Myrs before the end of our simulation, and there simply is not enough time for the heightened galactic perturbations to erode the outer Oort Cloud and conform to the correlation seen in our other solar analogs.  We also note that the median trapping efficiency for our control cases is 2.91\%, which is greater than all but 7 of our 31 solar analogs.  However, the other two control runs have trapping efficiencies of 2.27\% and 3.37\%, suggesting that the control cases have a similar level of scatter as the trend shown in Figure \ref{fig:tmin}.  

\subsubsection{Cloud Structure}

Although the Oort Cloud trapping efficiency is fairly independent of the Sun's galactocentric distance, this is not true of the distribution of orbits in the Oort Cloud.  If the Sun spends time significantly closer to the Galactic center the strong tidal field and stellar passages it encounters will torque bodies' perihelia out of the planetary region at smaller semimajor axes.  Because of this, the Oort Cloud's minimum semimajor axis is very sensitive to the minimum galactocentric distance of the Sun during our 4-Gyr simulations.  In addition, the maximum semimajor axis of the Oort Cloud is also affected by the Sun's orbital history.  Oort Cloud bodies can be directly stripped from the Sun's potential if it moves closer to the Galactic center.  In addition, more comets are ejected from the solar system via injections into the planetary region.  Both of these effects increase the erosion of the outer Oort Cloud.  Hence, both the inner and outer edge of the Oort Cloud are sensitive to the Sun's migration within the galaxy.

\begin{figure}[tbp]
\centering
\includegraphics[scale=.87]{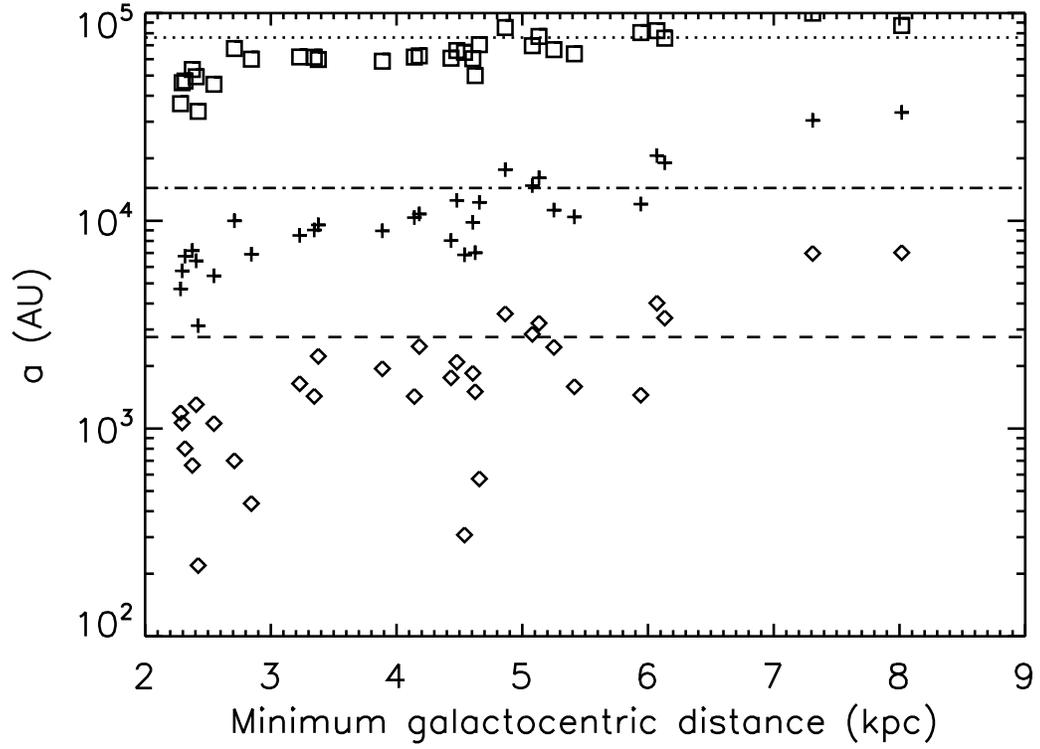}
\caption{Plot of $a_{min}$ ({\it diamonds}), $a_{max}$ ({\it squares}), and median semimajor axis ({\it crosses}) vs. the minimum galactocentric distance for each of our solar analogs.  In addition, we show $a_{min}$ ({\it dashed line}), $a_{max}$ ({\it dotted line}) and the median $a$ ({\it dash-dot-dash line}) for our control simulation.}\label{fig:7}
\end{figure}

In Figure \ref{fig:7}, we plot the minimum ($a_{min}$), maximum ($a_{max}$), and median semimajor axes for each of the Oort Clouds we have formed.  We define $a_{min}$ as the semimajor axis outside of which 95\% of all Oort Cloud bodies orbit.  Similarly, $a_{max}$ is defined as the semimajor axis inside of which 95\% of all Oort Cloud bodies orbit.  As can be seen in Figure \ref{fig:7}, all three semimajor axis parameters increase with increasing distance from the Galactic center.  Comparing our solar analogs to our control simulations, we see that for each semimajor axis parameter $\sim$75\% of our solar analogs have values below the control cases, in some cases by over an order of magnitude.  This is due to both the heightened erosion of the outer Oort Cloud and the heightened trapping in the extreme inner Oort Cloud that occurs at lower galactocentric distances.  Thus, Oort Cloud formation models that do not account for stellar migration tend to underestimate the cloud's central concentration.  

Although all Oort Cloud semimajor axes are correlated with the minimum galactocentric distance of the Sun, we see that there is a much greater amount of scatter around $a_{min}$ than $a_{max}$.  This is due to the stronger influence that stellar encounters have on the inner Oort Cloud's structure compared to the outer cloud, because the smallest semimajor axis orbits become trapped in the Oort Cloud due to the most powerful stellar encounters and not as a result of the Galactic tide.  Consequently, $a_{min}$ is influenced by a handful of isolated events, so there is a large amount of stochasticity in this parameter.  In contrast, scattered bodies can reach the outer cloud during any time period, regardless of stellar encounters.   Due to these differing effects, $a_{min}$ can vary by over an order of magnitude (between 200 and 7000 AU), while $a_{max}$ varies by less than a half decade between 30000 AU and 10$^{5}$ AU.  

\subsubsection{Cloud Isotropy}

\begin{figure}[tbp]
\centering
\includegraphics[scale=.87]{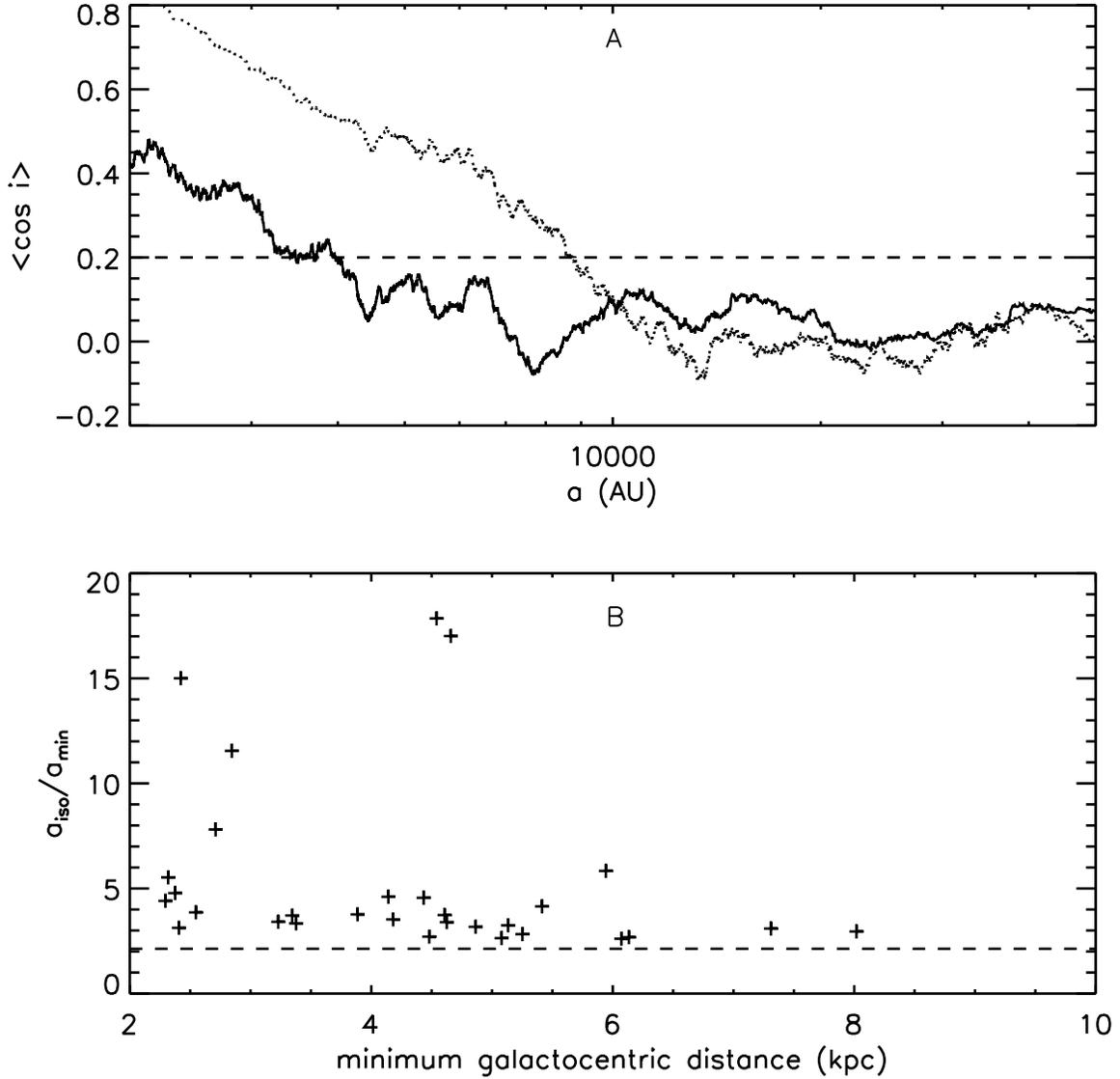}
\caption{{\it \bf a:} Plot of the mean cosine of the orbital inclination (relative to ecliptic) for two Oort Cloud simulations: the control simulation ({\it dotted line}) and Run 31 ({\it black line}).  The mean cosine is calculated by taking a moving average for the nearest 5\% of the Oort Cloud at a given semimajor axis.  The dotted line marks a mean cosine of 0.2, below which we consider the Oort Cloud to be isotropic.  {\it \bf b:} Plot of the ratio of the isotropization semimajor axis to the minimum semimajor axis of our Oort Clouds as a function of the minumum galactocentric distance attained by each solar analog.  The dotted line marks this semimajor axis ratio for our control simulation.}\label{fig:isotropy}
\end{figure}

Another prediction of previous simulations of Oort Cloud formation \citep{dones04, dun87} is that the isotropy of Oort Cloud orbits should vary with semimajor axis.  Because the Kozai tidal cycle \citep{heistre86} is longer than the age of the solar system at semimajor axes below $\sim$5000 AU and only a small fraction of stellar encounters strongly perturb such orbits, the innermost region of the Oort Cloud should have a distribution of orbital inclinations (relative to the ecliptic) still concentrated toward the ecliptic.  Indeed, when we look at the distribution of orbital inclinations as a function of semimajor axis for our control simulation in Figure \ref{fig:isotropy}a, we see that only beyond $a >$ 9000 AU is the mean cosine of inclination near 0 (which indicates isotropy).  However, we also plot the inclination distribution for Run 31 in Figure \ref{fig:isotropy}a.  As can be seen in this plot, the region of isotropic inclinations extends much deeper into the Oort Cloud than in our control run.  This is of course because when the effects of stellar migration are included, the strength of the local tidal field is not static and, in the case of Run 31, is larger than our control simulation for much of the history of the solar analog.  In addition, the numerical tidal field we measure in our simulated galaxy is not always as well-behaved as the analytical tides used in most previous works (see Section \ref{sec:galmeth}).  This element of stochasticity will also tend to increase the level of isotropy in the Oort Cloud.  

In Figure \ref{fig:isotropy}b, we plot the ratio of $a_{iso}$ to $a_{min}$ vs. the minimum galactocentric distance for each solar analog.  In addition, we also mark the median ratio for our control simulations, which at 2.31 is lower than any of our solar analogs.   For comparison, the vast majority of our simulations that include stellar migration have ratios that range from 2.5 to 6, with a few larger than 15.  Presumably, this is due to the fact once again that the strength of external perturbations vary much more in our simulations with stellar migration.  As a result, when the Sun is close to the Galactic center the stronger perturbing forces torque the perihelia of small orbits out of the planetary region and lower $a_{min}$.  However, the timescale to isotropize a distribution of inclinations is much longer than the timescale to shift a perihelion by tens of AU, so $a_{iso}$ is not decreased as much during these relatively short-lived excursions to low galactocentric distances.  In addition, the solar analogs with extremely high values of $a_{iso}/a_{min}$ represent cases that experienced exceptionally strong stellar encounters for the mean density of matter they encountered during their 4-Gyr history.  In these cases, this powerful stellar encounter torqued orbits into the Oort Cloud very deep in the scattered disk, but this perturbation was far too short to substantially affect the isotropy of the entire cloud.  

Although we have taken $a_{min}$ to represent the inner edge of the Oort Cloud for each of our solar analogs in the previous analyses, it is important to keep in mind that 5\% of our Oort Cloud bodies still orbit inside this value.  It is typically in this inner 5\% where we find the Sedna analogs that we discussed previously.

\subsubsection{Oort Cloud Erosion}

In the preceding sections, we have shown that solar analogs that spend time closer to the Galactic center will have more compact Oort Clouds than predicted by models that do not account for stellar migration.  On the other hand, the total amount of material that is trapped in the Oort Cloud does not increase if the Sun spends time at smaller galactocentric distances.  Because there is no enhancement in trapping efficiency, this means that the fraction of Oort Cloud bodies orbiting in the outer Oort Cloud ($a>$ 20,000 AU) will be smaller for solar analogs that spend an appreciable amount of time at small galactocentric distances.  

\begin{figure}[tbp]
\centering
\includegraphics[scale=.87]{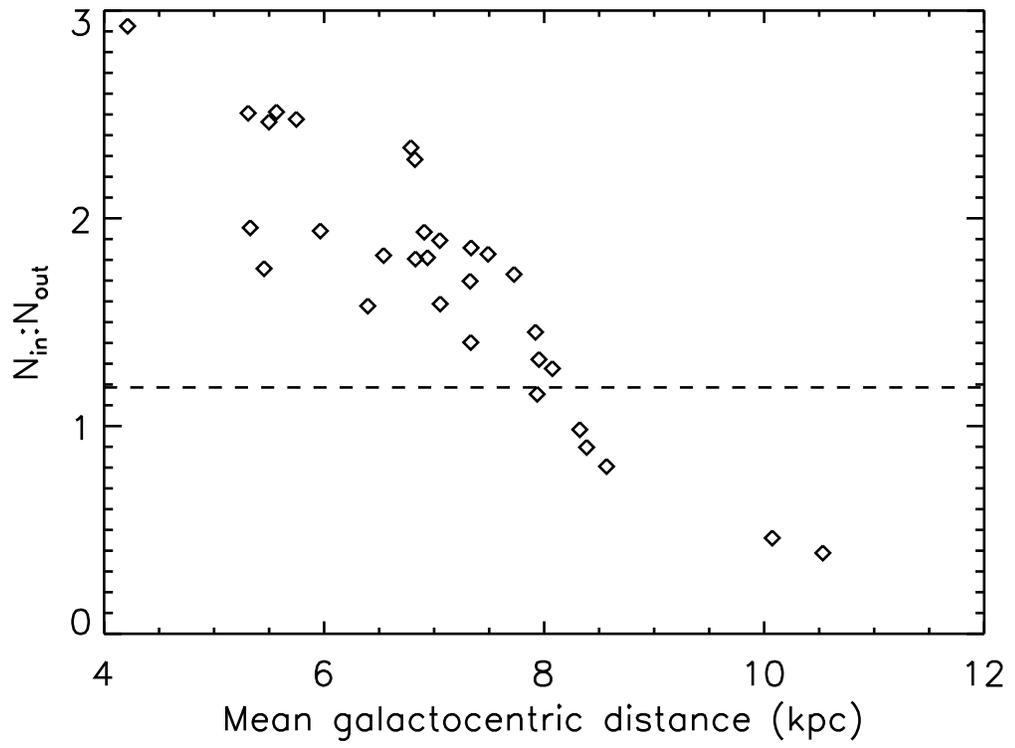}
\caption{{\it {\bf a:}}  Plot of the ratio of Oort Cloud bodies with 5,000 AU $<a<$ 20,000 AU ($N_{in}$) to bodies with $a>$ 20,000 AU ($N_{out}$) vs. the mean galactocentric distance for each solar analog.  The dashed line is the median fraction for our 3 control runs.}\label{fig:OCfrac}
\end{figure}

To demonstrate this, for each of our solar analogs we look at the population ratio of bodies orbiting between 5,000 AU $<a<$ 20,000 AU to bodies orbiting beyond 20,000 AU.  (For the purposes of the following discussion, we are only interested in bodies capable of evolving to observable comets, so we only consider bodies beyond $a=$ 5,000 AU \citep{kaibquinn09}.)  This ratio is plotted in Figure \ref{fig:OCfrac}.  We see that this ratio varies by nearly an order of magnitude depending on the Sun's orbital history and has a median value of 1.80, indicating that most comets typically orbit inside $a=$ 20,000 AU for our solar analogs.  For comparison, our control simulations predict a more evenly split distribution of comets with a median value of 1.19.  Looking again at Figure \ref{fig:7}, we see that the inner edge of the Oort Cloud is inside $a=$ 5,000 AU for all but two of our solar analogs.  Therefore, we can conclude that the trend in Figure \ref{fig:OCfrac} is due to the enhanced erosion of the outer Oort Cloud rather than the populating of low semimajor axes orbits.  

This presents additional difficulties for traditional models of Oort Cloud formation and comet delivery.  It is well-documented that the mass of the outer Oort Cloud implied by the observed long-period comet (LPC) flux near Earth seems to be at odds with other aspects of solar system formation \citep{dones04}.  If the outer Oort Cloud is the source of all LPCs there must be at least $\sim$4 x 10$^{11}$ comet-sized bodies orbiting in the outer Oort Cloud to produce the observed flux \citep{fran05}.  Assuming a typical comet mass to be $\sim$4 x 10$^{16}$ g \citep{weiss96}, we find that there should be roughly 2-3 M$_{\earth}$ of material in the outer Oort Cloud.  Our current control simulations as well as similar simulations in previous works \citep{kaibquinn08, dones04} indicate that only 1-3\% of all bodies scattered by the giant planets are trapped in the present day outer Oort Cloud.  Thus, to build the outer Oort Cloud seems to require an original disk of planetesimals with a mass of 70--300 M$_{\earth}$.  In contrast, a disk of no more than about 50 M$_{\earth}$ is favored by models attempting to explain the orbital evolution of the giant planets and trans-Neptunian objects \citep{morb07}.  This mass discrepancy typically becomes even worse when the enhanced erosion due to solar migration is included. 

Two possible solutions have recently emerged for this dilemma.  First, it has been shown that the outer Oort Cloud may not be the only or even the dominant source of observable LPCs \citep{kaibquinn09}.  Integrations of particles with semimajor axes between 5,000 and 20,000 AU indicate that they too can evolve to become LPCs via a dynamical pathway that is only moderately less efficient than outer Oort Cloud comet production.  Aside from expanding the semimajor axis range of comet production, this may also help solve this mass discrepancy because the trapping efficiency of the inner Oort Cloud can be an order of magnitude larger than that of the outer cloud if Oort Cloud formation began within the solar birth cluster \citep{bras06, kaibquinn08}.  Thus, the required protoplanetary disk mass to account for today's comet flux could be lower by a factor of 5-10 than if we only considered the outer Oort Cloud as an LPC source.  (In order to isolate the effects of stellar migration, we have no simulations that include the effects of a birth cluster.)


A second possible mechanism to ease the mass requirements of the protoplanetary disk was recently proposed in \citet{lev10}.  In this scenario, the Oort Cloud is not primarily comprised of material native to the solar system.  Instead, as the solar birth cluster dissolves, small bodies that were ejected from the planetary systems of other cluster members are captured into the Sun's potential well.  If the Oort Cloud formed by this mechanism, nearly all of the material in the cloud was deposited when the solar birth cluster dispersed, which was likely no more than a few Myrs after the Sun formed.  A second advantage of this model is that it is capable of producing a ratio of scattered disk to Oort Cloud bodies that is more consistent with that observed in the actual solar system \citep{dones04, dunlev97}.  However, for this formation mechanism to work, a significant fraction of the bodies in the Oort Cloud must be able to survive for the 4.5 Gyrs leading up to the present epoch.

To determine whether a captured Oort Cloud can survive for 4.5 Gyrs, we examine the fates of all particles that were added to our solar analogs' clouds within the first Gyr of our simulation.  As these particles evolve for the next 3 Gyrs, we calculate $f_{loss}$, the fraction of these bodies eroded per Gyr during this time period.  Assuming that all of these particles were instead added to the Oort Cloud at $t=0$, we can predict the fraction that will remain after 4.5 Gyrs of evolution with 

\begin{equation}
f_{remain} = (1-f_{loss})^{4.5}
\end{equation}

\begin{figure}[tbp]
\centering
\includegraphics[scale=.87]{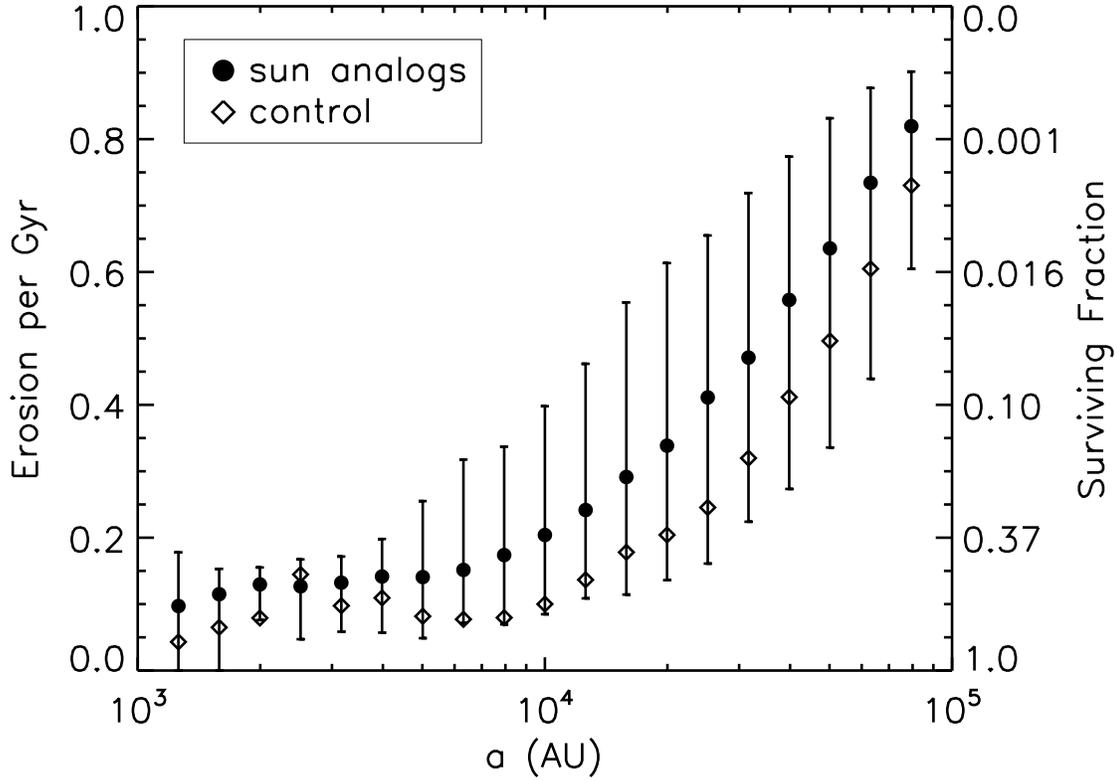}
\caption{Plot of the fraction of the Oort Cloud population eroded per Gyr as a function of the initial semimajor axes that particles have when they enter the Oort Cloud.  Semimajor axes bins are 0.2 dex wide in $\log{a}$-space.  Dots mark the median erosion value in each bin, while the error bars mark the range between the lowest 10\% of erosion rates and the highest 10\% of erosion rates.  Diamonds mark the median erosion rates for our control runs.  The righthand y-axis marks the projected population fraction that will remain if one assumes the erosion rates of the lefthand y-axis.}\label{fig:fracvsa}
\end{figure}

For each of our solar analogs, we calculate both the erosion rate and the predicted remaining fraction as a function of particles' initial Oort Cloud semimajor axis.  These results are displayed in Figure \ref{fig:fracvsa}.  We see that for the median case of our solar analogs, any Oort Cloud population captured beyond $a\sim$ 20,000 AU will be eroded by over an order of magnitude after 4.5 Gyrs of evolution, negating much of the population gains offered by the capture model in this $a$-range.  However, the situation improves inside 20,000 AU.  In this semimajor axis range, between 20 and 50\% of the population will survive for 4.5 Gyrs for our median case.  However, these captured bodies must be able to explain today's LPC flux, so they must be captured on semimajor axes beyond $\sim$5,000 AU \citep{kaibquinn09}.  Thus, it seems that non-native comets must be captured on orbits with initial semimajor axes between $\sim$5,000 AU and $\sim$20,000 AU in order for a large fraction to survive and account for today's observed LPCs.  

\section{Conclusions}
\label{sec:galcon}

In a simulation of a Milky Way analog, we have isolated stars at the end of the simulation with solar-like ages, galactic positions, and kinematics.  We find that the galactocentric distances of most of these stars has varied dramatically during their histories, and there appear to be many different orbital histories possible for stars like the Sun.  On the whole, however, our simulation indicates that it is possible that the Sun formed at least 2 kpc closer to the Galactic center than where it currently resides.  A large fraction of the solar analogs in our simulation spent at least a few hundred Myrs orbiting within 5 kpc or less of the Galactic center.  Past orbits that take the Sun within 2.5 kpc of the Galactic center are even possible.  Because the Sun would have encountered much higher disk densities at these small galactocentric distances, the stellar encounters and galactic tides that sculpt the structure of the Oort Cloud would likely have been stronger in the past.  

This has important consequences for our understanding of the orbit of Sedna.  This object in unaffected by the current Galactic tide, and therefore it has been assumed that it cannot belong to the conventional Oort Cloud.  Our work shows that this may not be the case.  Contrary to previous understanding, the minimum semimajor axis of orbits whose perihelia can be torqued beyond the planets is very dependent on the few most powerful field stellar passages, which are stochastic events.  In addition, we show that it is quite possible that the Sun spent much of its history residing in denser regions of the Milky Way, enhancing the chance for a ``Sedna-perturbing" stellar encounter.  The combination of these two findings suggests that Sedna may in fact represent the extreme inner edge of the conventional Oort Cloud and may be a result of the Sun's migration within the Milky Way.  Out of all of the solar analogs we study, $\sim$20-30\% generated Sedna analogs.  While Sedna is not the median case, our mechanism is certainly a plausible explanation for this object's orbit.

Most previous Oort Cloud simulations do not consider changes in the Galactic tide or the population of passing field stars, and we find that changes induced by the Sun's orbital history can have significant effects on the formation of the entire Oort Cloud.  When the Sun encounters stronger tidal and stellar perturbations the rate of perihelion change for orbits of any semimajor axis will increase.  Consequently, smaller semimajor axis orbits will be torqued out of the planetary region and into the Oort Cloud.  In addition, distant Oort Cloud bodies will be stripped or cycled back into the planetary region (and subsequently ejected) at a faster rate when the Sun has a small galactocentric distance.  Both of these effects cause the minimum and median orbital semimajor axis in the whole Oort Cloud to vary about an order of magnitude depending on the Sun's orbital history.  In general, an Oort Cloud formation model that assumes a constant Galactic tide overestimates the semimajor axes of Oort Cloud orbits compared to most models that include solar orbital migration throughout the galaxy. 

While the orbital structure of the Oort Cloud is sensitive to the Sun's past minimum galactocentric distance, we find that the fraction of scattered bodies that become trapped in the cloud is not correlated well with this parameter and range between 1.0 and 3.5\% .  This is because the timing of the Sun's period of lowest galactocentric distance determines whether the stronger galactic perturbations will have an erosive or an enriching effect on the Oort Cloud.  As stated already, a stronger tidal force will increase the efficiency at which smaller semimajor axis orbits are transferred from the scattered disk to the Oort Cloud.  However, the population of the scattered disk decreases throughout the history of the solar system, and if the scattered disk is largely depleted, a period of stronger perturbations will do little to enrich the Oort Cloud.  On the other hand, there will always be an increased erosion of the outer Oort Cloud.  For this reason, stars that reach their minimum galactocentric distances late in their histories tend to have the smallest Oort Cloud populations.

The increased cloud erosion experienced by most solar analogs also has implications for the source of observed long-period comets.  Historically, the outer Oort Cloud has been considered the source of LPCs.  However, only a small fraction of bodies scattered by the giant planets are trapped in the outer Oort Cloud, and an uncomfortably massive protoplanetary disk is required to form an outer Oort Cloud capable of generating current LPCs.  We show that solar migration within the Milky Way typically enhances the outer Oort Cloud's erosion, worsening this mass discrepancy.  On the other hand, the inner Oort Cloud is a more protected reservoir, strengthening the case that it may actually supply a significant fraction of observed LPCs \citep{kaibquinn09, dybkro11}.  

Lastly, this heightened erosion also has implications for alternative models of Oort Cloud formation.  Recently, it has been argued that the population of the Oort Cloud has been greatly bolstered by the capture of non-native bodies early in the solar system's history \citep{lev10}.  However, these bodies must be able to survive for 4.5 Gyrs to explain the modern LPC flux.  We find that this captured population will typically be eroded by at least an order of magnitude beyond $a\sim$ 20,000 AU.  Thus, bodies captured into inner Oort Cloud orbits ($a<$ 20,000 AU) show the most promise of accounting for the Oort Cloud's large population.

\section{Acknowledgements}
We would like to thank the two anonymous referees whose comments and suggestions greatly improved the quality of this paper.  This work was funded by a NASA Earth and Space Science Fellowship, a CITA National Fellowship, the National Science Foundation (grant AST-0709191) and NSERC.  The bulk of our computing was performed using Teragrid resources at Purdue University managed with Condor scheduling software (\url{http://www.cs.wisc.edu/condor}).  We also thank Martin Duncan and Mark Claire for helpful discussions.  

\clearpage

\begin{table}[htbp]
\centering
\begin{tabular}{c c c c c c c}
\hline
Galaxy & $v_{rot}$ & $\sigma_{tot}$ & $r_L$ & $r_H$ & $\rho$ & $\rho_*$\\ 
 & (km/s) & (km/s) & (kpc) & (pc) & (M$_{\sun}$/pc$^3$ & (M$_{\sun}$/pc$^3$) \\[0.5ex]
\hline
Milky Way & 220 & 42\footnotemark & 2.6 & 300 & 0.10 & 0.04\\
Simulated & 240 & 64 & 2.3--3.5\footnotemark & 300 & 0.09 & 0.056\\
\hline
\end{tabular}
\caption{Comparison between the properties of the Milky Way and our simulated galaxy at $t = 10$ Gyrs.  Columns are (from left to right): galaxy name, circular velocity at $r = 8.0$ kpc, stellar velocity dispersion at $r = 8.0$ kpc, disk scale length, disk scale height, total disk midplane density at $r = 8.0$ kpc, and stellar disk midplane density at $r = 8.0$ kpc.  Milky Way disk scale length and height are taken from \citet{jur08}, while the circular velocity is based on \citet{hou09}.  The Milky Way's total midplane and stellar densities are from \citet{holmflyn00} and \citet{justjahr10} respectively.  Finally, the stellar dispersion is based on \citet{garc01}. }
\label{tab:gal}
\end{table}

\footnotetext[1]{We chose to report the local Milky Way's dispersion value from \citet{garc01} since this work supplies the mass-velocity relation used in our stellar encounter code.  This dispersion value is based on Hipparcos velocity data of the solar neighborhood.  However, more recent observations indicate a higher local dispersion value of $\sim65$ km/s, which is much closer to our simulated galaxy \citep{holm09}.}
\footnotetext[2]{The scale length of our galaxy measured via surface density is 3.5 kpc, but the midplane density has a substantially smaller scale length because our disk's scale height grows with galactic radius.  Depending on the method of measurement, we find a scale length as small as 2.3 kpc.}

\clearpage

\begin{table}[tbp]
\centering
\begin{tabular}{c c c c c c c}
\hline
stars & age & $r$ & $z$ & $u$ & $v$ & $w$ \\ 
 & (Gyrs) & (kpc) & (pc) & (km/s) & (km/s) & (km/s) \\[0.5ex]
\hline
Sun & 4.57 & 8.4 $\pm$ 0.6 & 25 & 10 & 18 $\pm$ 2 & 7.2 \\
Sun Analogs & 4.45 $\pm$ 0.15 & 8.05 $\pm$ 0.7 & 0 $\pm$ 100 & 0 $\pm$ 15 & 0 $\pm$ 15 & 0 $\pm$ 15 \\
\hline
\end{tabular}
\caption{Comparison between the kinematical properties of the Sun and our selection criteria for solar analogs in our simulated galaxy at $t = 10$ Gyrs.  Column are (from left to right): star category name, stellar age, distance from galactic center, distance from the disk midplane, radial peculiar velocity, tangential peculiar velocity, and vertical peculiar velocity.  Solar distances to the Galactic center and above the midplane are taken from \citet{reid09} and \citet{jur08} respectively.  Solar velocity data are taken from \citet{macbin10}.}
\label{tab:sun}
\end{table}

\clearpage

\begin{longtable}{c c c c c c c c c c c}
\hline
Run & $r_{form}$ & $r_{min}$ & $r_{max}$ & $t(r_{min})$ & $a_{sed}$ & $a_{min}$ & $a_{med}$ & $a_{max}$ & $a_{iso}$ & Trap \%\\ 
 & (kpc) & (kpc) & (kpc) & (Gyr) & (AU) & (AU) & (AU) & (AU) & (AU) &\\[0.5ex]
\hline
\endhead
controla &   - &   - &   - &   - & 1184 & 2740 & 14382 & 76278 & 8729 & 2.91\\
controlb &   - &   - &   - &   - & 1177 & 2795 & 15338 & 75051 & 11395 & 3.37\\
controlc &   - &   - &   - &   - & 780 & 2758 & 13870 & 76627 & 7145 & 2.27\\
   1 &  5.48 &  4.18 &  9.77 & 0.10 & 1450 & 2485 & 10788 & 62242 &  8769 & 2.83\\
   2 &  8.37 &  6.07 & 10.43 & 4.14 & 1267 & 4017 & 20596 & 82286 & 10487 & 2.37\\
   3 &  8.21 &  4.66 & 11.03 & 2.68 & 614 & 573 & 12266 & 70461 &  9750 & 1.39\\
   4 &  4.18 &  2.38 &  7.97 & 3.40 & 495 & 665 &  7184 & 53492 &  3180 & 1.93\\
   5 &  3.89 &  3.89 &  8.99 & 0.00 & 1056 &1943 &  8946 & 58641 &  7321 & 3.30\\
   6 &  8.09 &  8.02 & 12.87 & 4.05 & 2789 &7025 & 33198 & 87156 & 20739 & 3.59\\
   7 &  7.08 &  2.71 &  9.17 & 3.72 &  580 &700 & 10005 & 67510 &  5466 & 1.70\\
   8 &  8.49 &  2.84 &  9.74 & 3.41 & 475 & 436 &  6899 & 60000 &  5035 & 1.03\\
   9 &  3.40 &  2.32 &  8.94 & 3.71 & 390 & 802 &  6757 & 47102 &  4434 & 1.50\\
  10 &  3.19 &  2.42 &  8.70 & 2.18 & 213 & 219 &  3126 & 33658 &  3290 & 2.68\\
  11 &  5.99 &  3.23 &  8.88 & 3.27 & 794 & 1645 &  8503 & 61498 &  5615 & 1.91\\
  12 &  3.92 &  2.29 &  9.61 & 0.04 & 688 & 1053 &  5708 & 45630 &  4664 & 2.58\\
  13 &  5.74 &  2.55 &  8.46 & 3.23 & 541 & 1058 &  5432 & 45403 &  4086 & 1.72\\
  14 &  5.56 &  4.61 & 10.29 & 0.28 & 1031 & 1846 &  9831 & 60306 &  6895 & 3.44\\
  15 &  8.44 &  6.13 & 10.76 & 2.65 & 1382 & 3408 & 19026 & 75725 &  9134 & 2.91\\
  16 &  7.50 &  4.48 & 10.26 & 2.40 & 1133 & 2089 & 12535 & 65991 &  5659 & 2.25\\
  17 &  5.90 &  4.14 & 10.11 & 0.08 & 920 & 1428 & 10366 & 61410 &  6577 & 2.67\\
  18 &  8.16 &  5.94 &  9.86 & 4.33 & 925 & 1449 & 12029 & 80551 &  8460 & 1.96\\
  19 &  7.99 &  4.86 & 11.48 & 3.97 & 1657 & 3568 & 17624 & 85213 & 11332 & 2.93\\
  20 &  7.51 &  7.31 & 13.42 & 0.02 & 2248 & 6972 & 30412 & 99913 & 21551 & 2.98\\
  21 &  5.45 &  2.41 &  8.78 & 3.60 & 679 & 1306 &  6414 & 49349 &  4083 & 1.72\\
  22 &  8.02 &  4.43 &  8.76 & 3.35 & 762 & 1758 &  8032 & 60661 &  8009 & 1.03\\
  23 &  5.75 &  4.54 &  9.99 & 2.09 & 254 & 307 &  6856 & 64671 &  5489 & 1.49\\
  24 &  5.87 &  3.38 & 10.62 & 0.48 & 1406 & 2229 &  9552 & 59767 &  7445 & 2.68\\
  25 &  6.68 &  5.41 &  8.81 & 1.73 & 817 & 1591 & 10436 & 63668 &  6597 & 2.18\\
  26 &  5.43 &  4.63 &  9.20 & 2.44 & 938 & 1506 &  7016 & 50018 &  5114 & 1.85\\
  27 &  8.97 &  5.13 & 10.36 & 2.65 & 1443 & 3215 & 16102 & 77206 & 10434 & 1.69\\
  28 &  5.70 &  5.25 &  9.92 & 2.34 & 1145 & 2464 & 11259 & 66689 &  6968 & 2.63\\
  29 &  6.21 &  5.08 & 10.38 & 0.05 & 908 & 2845 & 14774 & 69655 &  7503 & 3.40\\
  30 &  6.91 &  3.35 &  9.29 & 1.63 & 719 & 1431 &  9033 & 61311 &  5317 & 2.10\\
  31 &  5.10 &  2.28 &  8.51 & 1.45 & 685 & 1190 &  4701 & 36612 &  3416 & 2.66\\
\hline
\caption{Summary of Oort Cloud properties for each of our simulations.  The columns from left to right are: Simulation name, distance at which the solar analog formed, minimum galactocentric distance attained by the solar analog, maximum galactocentric distance attained by the solar analog, time at which solar analog attained minimum galactocentric distance, value of $a_{sed}$, minimum semimajor axis in Oort Cloud, median semimajor axis in Oort Cloud, maximum semimajor axis in Oort Cloud, Oort Cloud semimajor axis beyond which orbital inclinations are isotropized, percentage of total particles trapped in Oort Cloud after 4 Gyrs.}\\

\label{tab:runs}
\end{longtable}
\clearpage

\clearpage

\clearpage

\clearpage

\clearpage

\clearpage

\clearpage

\clearpage

\clearpage

\clearpage

\clearpage

\clearpage

\clearpage

\clearpage

\clearpage

\clearpage

\clearpage

\clearpage

\clearpage

\bibliographystyle{icarus}
\bibliography{OCmigrate}

\end{document}